\documentstyle[11pt,epsfig,twoside,cite,amssymb]{article}
\newcommand{\bc}{\begin{center}}
\newcommand{\ec}{\end{center}}
\begin{document}
\begin{flushright}
PRA-HEP 99-07
\end{flushright}
\begin{center}
{\Large \bf Interpreting virtual photon interactions in terms of
parton distribution functions}

\vspace*{0.7cm}
Ji\v{r}\'{\i} Ch\'{y}la and Marek Ta\v{s}evsk\'{y}

\vspace*{0.2cm}
{\em Institute of Physics, Na Slovance 2, Prague 8, Czech Republic}

\vspace*{0.5cm}
{\large \bf Abstract}
\begin{quotation}
\noindent
Interactions of virtual photons are analyzed in terms of photon
structure. It is argued that the concept of parton distribution
functions is phenomenologically very useful even for highly virtual
photons involved in hard collisions. This claim is illustrated on
leading order expressions for $F_2^{\gamma}(x,P^2,Q^2)$ and
effective parton distribution function $D_{\mathrm eff}(x,P^2,Q^2)$
relevant for jet production, as well as within the
next--to--leading order QCD calculations of jet
cross--sections in electron--proton collisions.
\end{quotation}
\end{center}

\section{Introduction}
Parton distribution functions (PDF) are, together with the colour
coupling $\alpha_s$, the basic ingredients of perturbative QCD
calculations. It is worth emphasizing that in quantum field theory
it is difficult to distinguish effects of the ``structure'' from
those of ``interactions''. Within the Standard Model (SM) it makes
good sense to distinguish {\em fundamental particles}, which
correspond to fields in its lagrangian ${\cal L}_{\mathrm{SM}}$
(leptons, quarks and gauge bosons) from {\em composite particles},
which appear in the mass spectrum but have no corresponding field
in ${\cal L}_{\mathrm{SM}}$. For the latter the use of PDF to
describe their ``structure'' appears natural, but the concept of
PDF turns out to be phenomenologically useful also for some
fundamental particles, in particular the photon. PDF are
indispensable for the real photon due to strong interactions
between the $q\overline{q}$ pair to which it couples
electromagnetically.
For massless quarks this coupling leads to singularities,
which must be absorbed in PDF of the photon, similarly as in the
case of hadrons. Although PDF of the real photon satisfy
inhomogeneous evolution equations, their physical meaning remains
basically the same as for hadrons.

For nonzero photon virtualities there is no true singularity
associated with the coupling $\gamma^*\rightarrow q\overline{q}$
and one therefore expects that for sufficiently virtual photon its
interactions should be calculable perturbatively, with no need
to introduce PDF. The main aim of this paper is to advocate the use
of PDF also for virtual photons involved in hard collisions.

Throughout this paper we shall stay within the conventional approach
to evolution equations for PDF of the photon. The reformulation of
the whole framework for the description of hard collisions involving
photons in the initial state, proposed recently
by one of us \cite{jfactor}, affects the analysis of dijet
production in photon--proton collisions, which is the main subject
of this paper, only marginally. This
stands in sharp contrast to QCD analysis of $F_2^{\gamma}$
which is affected by this reformulation quite significantly.

The paper is organized as follows. In Section 2 the notation and
basic facts concerning the evolution equations for PDF of the real
photon and the properties of their solutions are recalled.
In Section 3 the physical content of PDF of the virtual photon is
analyzed and its phenomenological relevance illustrated within the
LO QCD. This is further underlined in Section 4 by detailed analysis
of NLO QCD calculations of dijet cross--sections in ep collisions,
followed by the summary and conclusions in Section 5.

\section{PDF of the real photon}
In QCD the coupling of quarks and gluons is characterized by the
renormalized colour coupling (``couplant'' for short)
$\alpha_s(\mu)$, depending on the {\em renormalization scale} $\mu$
and satisfying the equation
\begin{equation}
\frac{{\mathrm d}\alpha_s(\mu)}{{\mathrm d}\ln \mu^2}\equiv
\beta(\alpha_s(\mu))=
-\frac{\beta_0}{4\pi}\alpha_s^2(\mu)-
\frac{\beta_1}{16\pi^2}
\alpha_s^3(\mu)+\cdots,
\label{RG}
\end{equation}
where, in QCD with $n_f$ massless quark flavours, the first two
coefficients, $\beta_0=11-2n_f/3$ and $\beta_1=102-38n_f/3$, are
unique, while all the higher order ones are ambiguous. As we shall
stay in this paper within the NLO QCD, only the first two, {\em
unique}, terms in (\ref{RG}) will be taken into account.
However, even for a given r.h.s. of (\ref{RG}) its
solution $\alpha_s(\mu)$ is not a unique function of $\mu$,
because there is an infinite number of solutions of (\ref{RG}),
differing by the initial condition. This so called {\em
renormalization scheme} (RS) ambiguity \footnote{In higher orders
this ambiguity includes also the arbitrariness of the coefficients
$\beta_i,i\ge 2$.}
can be parameterized in a number of ways. One
of them makes use of the fact that in the process of
renormalization another dimensional parameter, denoted usually
$\Lambda$, inevitably appears in the theory. This parameter
depends on the RS and at the NLO even fully specifies it.
For instance, $\alpha_s(\mu)$ in
the familiar MS and $\overline{\mathrm {MS}}$ RS are two
solutions of the same equation (\ref{RG}), associated with different
$\Lambda_{\mathrm {RS}}$. At the NLO the variation of both the
renormalization scale $\mu$ and the renormalization scheme
RS$\equiv$\{$\Lambda_{\mathrm {RS}}$\} is redundant. It
suffices to fix one of them and vary the other, but we stick
to the common practice of considering both of them as free
parameters. In this paper we shall work in the standard
$\overline{\mathrm {MS}}$ RS of the couplant.

The ``dressed''
\footnote{In the following the adjective ``dressed'' will be
dropped, and if not stated otherwise, all PDF will be understood to
pertain to the photon.} PDF
result from the resummation of multiple parton collinear emission
off the corresponding ``bare'' parton distributions. As a result of
this resummation PDF acquire dependence on the {\em factorization
scale} $M$. This scale defines the upper limit on
some measure $t$ of the off--shellness of partons included in the
definition of $D(x,M)$
\begin{equation}
D_i(x,M)\equiv \int_{t_{\mathrm {min}}}^{M^2}{\mathrm d}t
d_i(x,t),~~~~~~i=q,\overline{q},G,
\label{dressed}
\end{equation}
where the {\em unintegrated} PDF $d_i(x,t)$ describe distributions
of partons with the momentum fraction $x$ and {\em fixed}
off--shellness $t$. Parton virtuality $\tau\equiv\mid p^2-m^2\mid$
or transverse mass $m_T^2\equiv p_T^2+m^2$, are two standard
choices of such a measure.
Factorization scale dependence of PDF of the photon
is determined by the system of coupled inhomogeneous evolution
equations
\begin{eqnarray}
\frac{{\mathrm d}\Sigma(x,M)}{{\mathrm d}\ln M^2}& =&
\delta_{\Sigma}k_q+P_{qq}\otimes \Sigma+ P_{qG}\otimes G,
\label{Sigmaevolution}
\\ \frac{{\mathrm d}G(x,M)}{{\mathrm d}\ln M^2} & =& k_G+
P_{Gq}\otimes \Sigma+ P_{GG}\otimes G, \label{Gevolution} \\
\frac{{\mathrm d}q_{\mathrm {NS}}(x,M)}{{\mathrm d}\ln M^2}& =&
\delta_{\mathrm {NS}} k_q+P_{\mathrm {NS}}\otimes q_{\mathrm{NS}},
\label{NSevolution}
\end{eqnarray}
where
\begin{eqnarray}
\Sigma(x,M) & \equiv & \sum_{i=1}^{n_f}q_i^{+}(x,M)\equiv
\sum_{i=1}^{n_f} \left[q_i(x,M)+\overline{q}_i(x,M)\right],
\label{singlet}\\ q_{\mathrm{NS}}(x,M)& \equiv &
\sum_{i=1}^{n_f}\left(e_i^2-\langle e^2\rangle\right)
\left(q_i(x,M)+\overline{q}_i(x,M)\right),
\label{nonsinglet}
\end{eqnarray}
\begin{equation}
\delta_{\mathrm{NS}}=6n_f\left(\langle e^4\rangle-\langle
e^2\rangle ^2\right),~~~\delta_{\Sigma}=6n_f\langle e^2\rangle.
\label{sigmas}
\end{equation}
To order $\alpha$ the splitting functions $P_{ij}$ and
$k_i$ are given as power expansions in $\alpha_s(M)$:
\begin{eqnarray}
k_q(x,M) & = & \frac{\alpha}{2\pi}\left[k^{(0)}_q(x)+
\frac{\alpha_s(M)}{2\pi}k_q^{(1)}(x)+
\left(\frac{\alpha_s(M)}{2\pi}\right)^2k^{(2)}_q(x)+\cdots\right],
\label{splitquark} \\ k_G(x,M) & = &
\frac{\alpha}{2\pi}\left[~~~~~~~~~~~~
\frac{\alpha_s(M)}{2\pi}k_G^{(1)}(x)+
\left(\frac{\alpha_s(M)}{2\pi}\right)^2k^{(2)}_G(x)+\cdots\;\right],
\label{splitgluon} \\ P_{ij}(x,M) & = &
~~~~~~~~~~~~~~~~~~\frac{\alpha_s(M)}{2\pi}P^{(0)}_{ij}(x) +
\left(\frac{\alpha_s(M)}{2\pi}\right)^2 P_{ij}^{(1)}(x)+\cdots,
\label{splitpij}
\end{eqnarray}
where the leading order splitting functions
$k_q^{(0)}(x)=x^2+(1-x)^2$ and $P^{(0)}_{ij}(x)$ are {\em
unique}, while all higher order ones
$k^{(j)}_q,k^{(j)}_G,P^{(j)}_{kl},j\ge 1$ depend on the choice of
the {\em factorization scheme} (FS). The equations
(\ref{Sigmaevolution}-\ref{NSevolution}) can be rewritten as
evolution
equations for $q_i(x,M),\overline{q}_i(x,M)$ and $G(x,M)$ with
inhomegenous splitting functions $k_{q_i}^{(0)}=3e_i^2k_q^{(0)}$.
The photon structure function $F_2^{\gamma}(x,Q^2)$, measured in
deep inelastic scattering of electrons on photons is given as
a sum of convolutions
\begin{eqnarray}
\frac{1}{x}F_2^{\gamma}(x,Q^2)&=& q_{\mathrm{NS}}(M)\otimes
C_q(Q/M)+\frac{\alpha}{2\pi}\delta_{\mathrm{NS}}C_{\gamma}+
\label{NSpart} \\ & & \langle e^2\rangle \Sigma(M)\otimes
C_q(Q/M)+\frac{\alpha}{2\pi} \langle
e^2\rangle\delta_{\Sigma}C_{\gamma}+ \langle
e^2\rangle\frac{\alpha_s}{2\pi}G(M)\otimes C_G(Q/M)
\label{S+Gpart}
\end{eqnarray}
of photonic PDF and coefficient functions
$C_q(x),C_G(x),C_{\gamma}(x)$ admitting perturbative expansions
\begin{eqnarray}
C_q(x,Q/M) & = & \delta(1-x)~~~~+
~~~\frac{\alpha_s(\mu)}{2\pi}C^{(1)}_q(x, Q/M)+\cdots,
\label{cq} \\
C_G(x,Q/M) & = & ~~~~~~~~~~~~~~~~~~~~~
\frac{\alpha_s(\mu)}{2\pi}C^{(1)}_G(x,Q/M)+\cdots,
\label{cG} \\
C_{\gamma}(x,Q/M) & = &
C_{\gamma}^{(0)}(x,Q/M)+
\frac{\alpha_s(\mu)}{2\pi}C_{\gamma}^{(1)}(x,Q/M)+\cdots.
\label{cg}
\end{eqnarray}
The renormalization scale $\mu$, used as argument of
$\alpha_s(\mu)$ in (\ref{cq}-\ref{cg})
is in principle independent of the
factorization scale $M$. Note that despite the presence of $\mu$ as
argument of $\alpha_s(\mu)$ in (\ref{cq}--\ref{cg}), the
coefficient functions $C_q,C_G$ and $C_{\gamma}$, summed to all
orders of $\alpha_s$, are actually
independent of it, because the $\mu$--dependence of the
expansion parameter
$\alpha_s(\mu)$ is cancelled by explicit dependence of $C^{(i)}_q,
C^{(i)}_G,C^{(i)}_{\gamma},i\ge 2$ on $\mu$. On the
other hand, PDF as well as the coefficient functions $C_q, C_G$ and
$C_{\gamma}$ do depend on both the factorization scale $M$ and
factorization scheme, but in such a correlated manner that
physical quantities,
like $F_2^{\gamma}$, are independent of both $M$ and the FS,
provided expansions (\ref{splitquark}--\ref{splitpij}) and
(\ref{cq}--\ref{cg}) are taken to all orders in $\alpha_s(M)$ and
$\alpha_s(\mu)$. In practical calculations based on truncated forms
of (\ref{splitquark}--\ref{splitpij}) and (\ref{cq}--\ref{cg}) this
invariance is, however, lost and the choice of both $M$ and FS
makes numerical difference even for physical quantities. The
expressions for
$C_q^{(1)},C^{(1)}_G$ given in \cite{bardeen} are usually claimed
to correspond to ``$\overline{\mathrm {MS}}$ factorization scheme''.
As argued in \cite{jch2}, this denomination is, however,
incomplete. The adjective ``$\overline{\mathrm {MS}}$'' concerns
exclusively the choice of the RS of the couplant $\alpha_s$ and has
nothing to do with the choice of the splitting functions
$P^{(1)}_{ij}$. The choices of the
renormalization scheme of the couplant $\alpha_s$ and of the
factorization scheme of PDF are two completely independent
decisions, concerning two different and in general unrelated
redefinition procedures. Both are necessary in order to specify
uniquely the results of fixed order perturbative calculations, but
we may combine any choice of the RS of the couplant with any choice
of the FS of PDF. The coefficient functions
$C_q,C_G,C_{\gamma}$ depend on both of them,
whereas the splitting functions depend only on the latter. The
results given in \cite{bardeen} correspond to $\overline{\mathrm
{MS}}$ RS of the couplant but to the ``minimal subtraction'' FS of
PDF
\footnote{See Section 2.6 of \cite{FP}, in particular eq. (2.31),
for discussion of this point.}. It is therefore more appropriate to
call this full specification of the renormalization and factorization
schemes as ``$\overline{\mathrm {MS}}+{\mathrm {MS}}$ scheme''.
Although the phenomenological relevance of treating $\mu$ and $M$
as independent parameters has been demonstrated \cite{fontannaz},
we shall follow the usual practice of setting $\mu=M$.

\subsection{Pointlike solutions and their properties}
\begin{figure}\unitlength 1mm\centering
\begin{picture}(150,60)
\put(0,30){\epsfig{file=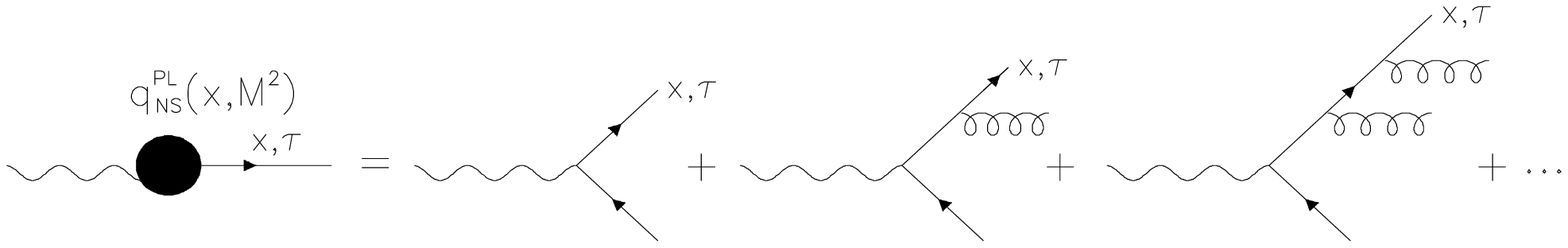,width=12cm}}
\put(0,0){\epsfig{file=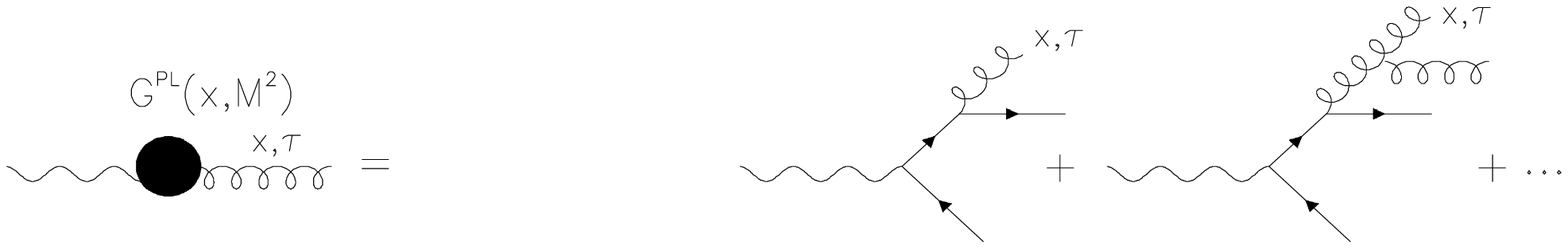,width=12cm}}
\end{picture}
\caption{Diagrams defining the pointlike parts of nonsinglet
quark and gluon distribution functions. The
resummation involves integration over parton virtualities $\tau\le
M^2$.}
\label{figpl}
\end{figure}

The general solution of the evolution equations
(\ref{Sigmaevolution}-\ref{NSevolution}) can be written as the sum
of a particular solution of the full inhomogeneous equation and the
general solution of the corresponding homogeneous one, called
{\em hadronic} \footnote{Sometimes also called ``VDM part''
because it is usually modelled by PDF of vector mesons.} part. A
subset of the solutions of full evolution equations resulting from
the resummation of series of diagrams like those
in Fig. \ref{figpl}, which
start with the pointlike purely QED vertex $\gamma\rightarrow
q\overline{q}$, are called {\em pointlike} (PL) solutions. In
writing down the expression for the resummation of diagrams in
Fig. \ref{figpl} there is a freedom in specifying some sort of
boundary condition. It is common to work within a subset of
pointlike solutions specified by the value of the scale $M_0$ at
which they vanish. In general, we can thus write
($D=q,\overline{q},G$)
\begin{equation}
D(x,M^2)= D^{\mathrm {PL}}(x,M^2)+D^{\mathrm{HAD}}(x,M^2).
\label{separation}
\end{equation}
Due to the fact that there is an infinite number of pointlike
solutions $q^{\mathrm{PL}}(x,M^2)$, the separation of quark and
gluon distribution functions into their pointlike and hadronic
parts is, however, ambiguous and therefore these concepts have
separately no physical meaning. In \cite{smarkem1} we
discussed numerical aspects of this ambiguity for the
Schuler--Sj\"{o}strand sets of parameterizations \cite{sas1}.

To see the most important feature of the
pointlike part of quark distribution functions that will be
crucial for the following analysis, let us consider in detail the
case of nonsinglet quark distribution function $q_{\mathrm
{NS}}(x,M)$, which is explicitly defined via the series
\begin{displaymath}
q^{\mathrm {PL}}_{\mathrm {NS}}(x,M_0,M) \equiv
\frac{\alpha}{2\pi}k_{\mathrm {NS}}^{(0)}(x)
\int^{M^2}_{M_0^2}\frac{{\mathrm d}\tau}{\tau}+
\int^{1}_{x}\frac{{\mathrm d}y}{y}P^{(0)}_{qq}
\left(\frac{x}{y}\right) \int^{M^2}_{M_0^2} \frac{{\mathrm
d}\tau_1}{\tau_1}\frac{\alpha_s(\tau_1)}{2\pi}\frac{\alpha}{2\pi}
k_{\mathrm {NS}}^{(0)}(y)\int^{\tau_1}_{M_0^2} \frac{{\mathrm
d}\tau_2}{\tau_2}+
\end{displaymath}
\begin{equation}
\int^{1}_{x}\frac{{\mathrm d}y}{y}P^{(0)}_{qq}
\left(\frac{x}{y}\right) \int^{1}_{y}\frac{{\mathrm
d}w}{w}P^{(0)}_{qq} \left(\frac{y}{w}\right) \int^{M^2}_{M_0^2}
\frac{{\mathrm d}\tau_1}{\tau_1}\frac{\alpha_s(\tau_1)}{2\pi}
\int^{\tau_1}_{M_0^2} \frac{{\mathrm
d}\tau_2}{\tau_2}\frac{\alpha_s(\tau_2)}{2\pi}\frac{\alpha}{2\pi}
k_{\mathrm {NS}}^{(0)}(w)\int^{\tau_2}_{M_0^2}\frac{{\mathrm
d}\tau_3}{\tau_3} +\cdots, \label{resummation}
\end{equation}
where $k_{\mathrm{NS}}^{(0)}(x)=\delta_{\mathrm{NS}}k^{(0)}_q(x)$.
In terms of moments defined as
\begin{equation}
f(n)\equiv \int_0^1x^n f(x)\mathrm{d}x
\label{moment}
\end{equation}
this series can be resummed in a closed form
\begin{equation}
q_{\mathrm {NS}}^{\mathrm {PL}}(n,M_0,M)=\frac{4\pi}{\alpha_s(M)}
\left[1-\left(\frac{\alpha_s(M)}{\alpha_s(M_0)}\right)^
{1-2P^{(0)}_{qq}(n)/\beta_0}\right]a_{\mathrm {NS}}(n),
\label{generalpointlike}
\end{equation}
where
\begin{equation}
a_{\mathrm {NS}}(n)\equiv \frac{\alpha}{2\pi\beta_0}
\frac{k_{\mathrm {NS}}^{(0)}(n)}{1-2P^{(0)}_{qq}(n)/\beta_0}.
\label{ans}
\end{equation}
\begin{figure}\centering
\epsfig{file=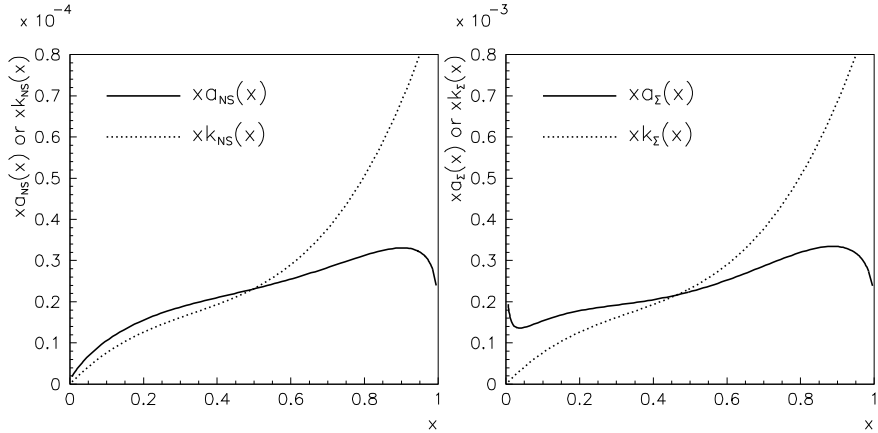,height=6cm}
\caption{
Comparison, left in the nonsinglet and right in the
singlet channels, of the functions
$xk_i(x)=x\delta_ik_q^{(0)},i=\mathrm{NS},\Sigma$ with the
function (\ref{ans}) and
its analogue in the singlet channel.}
\label{anskns}
\end{figure}
It is straightforward to show that
(\ref{resummation}) or, equivalently, (\ref{generalpointlike}),
satisfy the evolution equation (\ref{NSevolution}) with the
splitting functions $k_q$ and $P_{ij}$ including the first terms
$k_q^{(0)}$ and $P_{qq}^{(0)}$ only.

Transforming (\ref{ans}) to the $x$--space by means of inverse Mellin
transformation we get
$a_{\mathrm{NS}}(x)$ shown in Fig. 2. The resummation softens
the $x-$dependence of $a_{\mathrm {NS}}(x)$ with
respect to the first term in (\ref{resummation}), proportional to
$k_{\mathrm {NS}}(x)$, but does not change the logarithmic
dependence of $q_{\mathrm {NS}}$ on $M$. In the nonsinglet
channel the effects of gluon radiation on
$q_{\mathrm{NS}}^{\mathrm{PL}}$
are significant for $x>0.6$ but small elsewhere, whereas
in the singlet
channel such effects are marked also for $x<0.5$ and
 lead to a steep rise of $xq_{\mathrm{NS}}^{\mathrm{PL}}$
at very small $x$. As emphasized long time
ago by authors of \cite{FKP} the logarithmic dependence of
$q_{\mathrm{NS}}^{\mathrm{PL}}$ on $\ln M$ has
nothing to do with QCD and results exclusively from integration
over the transverse momenta (virtualities) of quarks coming from
the basic QED $\gamma^*\rightarrow q\overline{q}$ splitting.
For $M/M_0\rightarrow \infty$ the second term in brackets of
(\ref{generalpointlike}) vanishes and therefore all pointlike
solutions share the same large $M$ behaviour
\begin{equation}
q^{\mathrm {PL}}_{\mathrm {NS}}(x,M_0,M)
\rightarrow \frac{4\pi}{\alpha_s(M)}a_{\mathrm {NS}}(x)\equiv
q^{\mathrm {AP}}_{\mathrm {NS}}(x,M)\propto \ln \frac{M^2}{\Lambda^2},
\label{asymptotic}
\end{equation}
defining the so called {\em asymptotic pointlike} solution
$q^{\mathrm {AP}}_{\mathrm {NS}}(x,M)$ \cite{witten,BB}.
The fact that for
the asymptotic pointlike solution (\ref{asymptotic}) $\alpha_s(M)$
appears in the denominator of (\ref{asymptotic}) has been the
source of misleading claims that $q(x,M)=O(\alpha/\alpha_s)$.
In fact, as argued in detail in \cite{jfactor}, $q(x,M)=O(\alpha)$
as suggested by explicit construction in (\ref{resummation}).
We shall return to this point in Section 4.2 when discussing the
factorization scale invariance of finite order approximations
to dijet cross--sections.

The arbitrariness in the choice of $M_0$ reflects the fact that as
$M_0$ increases, less of the gluon radiation effects is included in
the resummation (\ref{generalpointlike}) defining the pointlike
part of quark distribution function $q^{\mathrm {PL}}$ but included
in hadronic one. The latter is usually modelled by the VDM ansatz
and will therefore be called ``VDM'' in the following. As we shall
see, the hadronic and pointlike parts have very different behaviour
as functions of $x$ and $M$.

\subsection{Properties of Schuler--Sj\"{o}strand parameterizations}
Practical aspects of the ambiguity in separating PDF into their VDM
and pointlike parts can be illustrated on the properties of SaS1D
and SaS2D parameterizations \cite{sas1}
\footnote{The properties of SaS1M and SaS2M parameterizations are
similar.}. What makes the SaS approach
particularly useful for our discussion is the fact that it provides
separate parameterizations of the VDM and pointlike parts
\footnote{Called for short ``pointlike quarks'' and ``pointlike
gluons'' in the following.}
of both quark and gluon distributions.
\begin{figure}[t]\centering
\epsfig{file=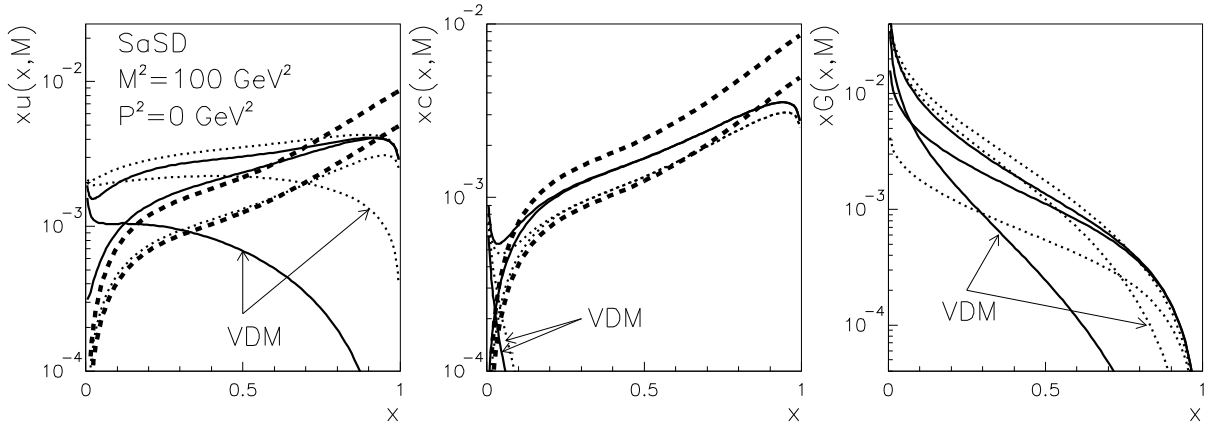,width=\textwidth}
\caption{The $u$ and $c$ quark and gluon distribution
functions of the real photon for SaS1D (upper solid curves) and
SaS2D (upper dotted curves) parameterizations at $M^2=100$ GeV$^2$.
The VDM and pointlike parts of both parameterizations are plotted
separately, the latter as solid and dotted curves peaking at large
$x$. For quarks the splitting terms (\ref{splitterm}) corresponding
to SaS1D and SaS2D (upper and lower dashed curves) are
overlayed to show the effects of resummation (\ref{resummation}).}
\label{sasd1real}
\end{figure}
\begin{figure} \unitlength 1mm
\begin{picture}(160,160)
\put(0,50){\epsfig{file=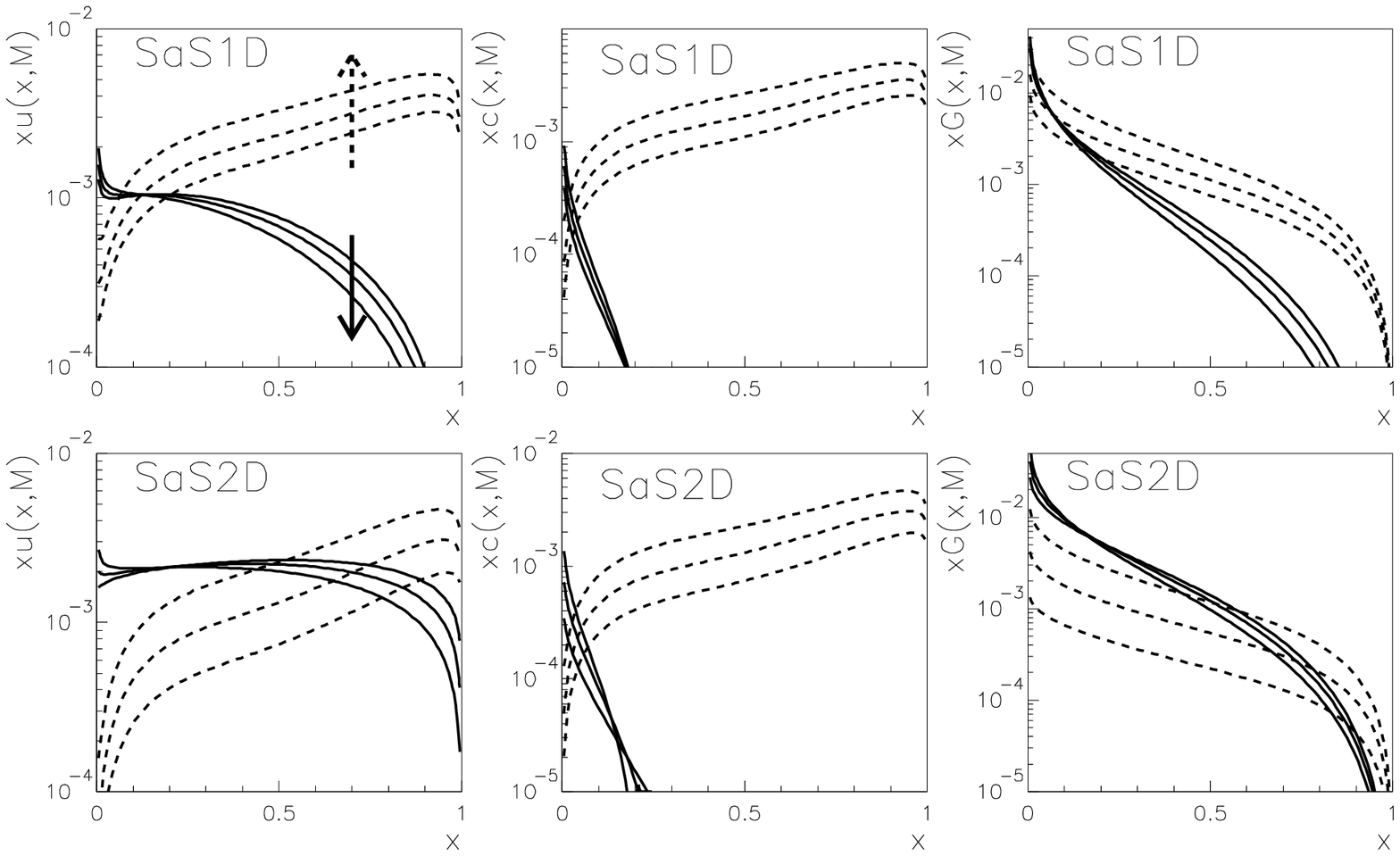,width=\textwidth}}
\put(0,0){\epsfig{file=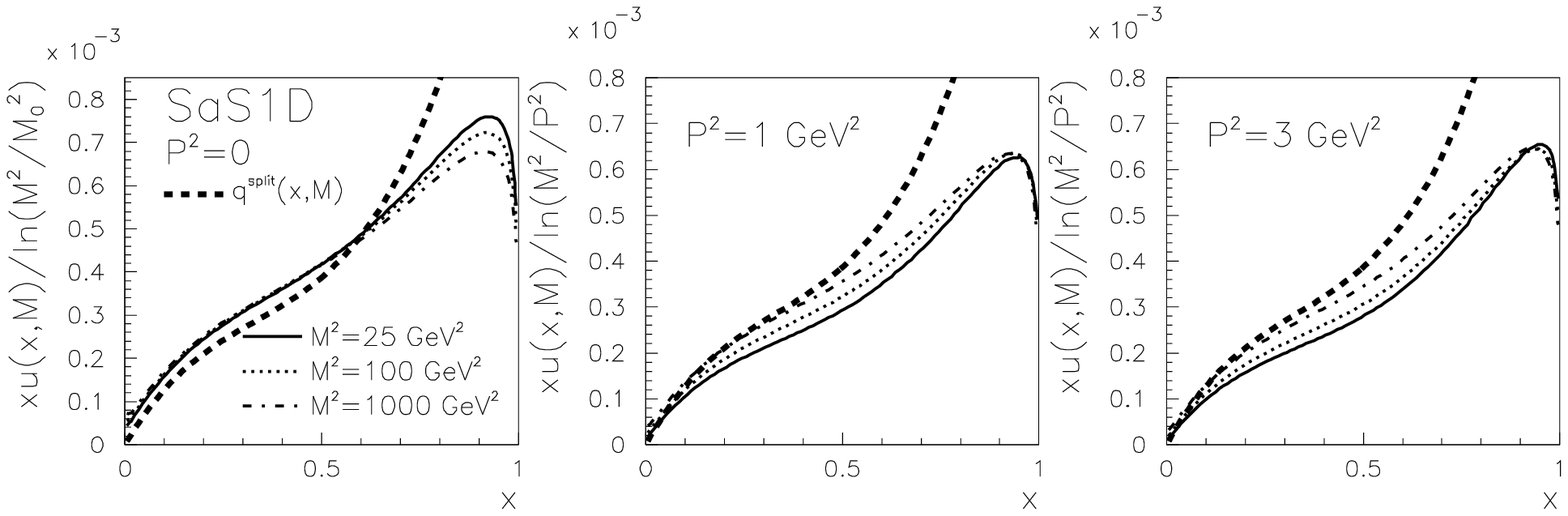,width=\textwidth}}
\end{picture}
\caption{Factorization scale dependence of parton distributions
functions $u(x,M),c(x,M)$ and $G(x,M)$ of the real photon. Dashed
and solid curves correspond, in the order indicated by the arrows,
to pointlike and VDM parts of these distributions at
$M^2=25,100$ and $1000$ GeV$^2$. In the lower part SaS1D
quark distribution functions $xu(x,M^2,P^2)$ rescaled by
$\ln(M^2/M_0^2)$ for the
real photon (left) and by $\ln(M^2/P^2)$ for the virtual one, are
plotted and compared to the predictions of the splitting term
(\ref{splitterm}).}
\label{scaledependence}
\end{figure}

\begin{figure}\centering
\epsfig{file=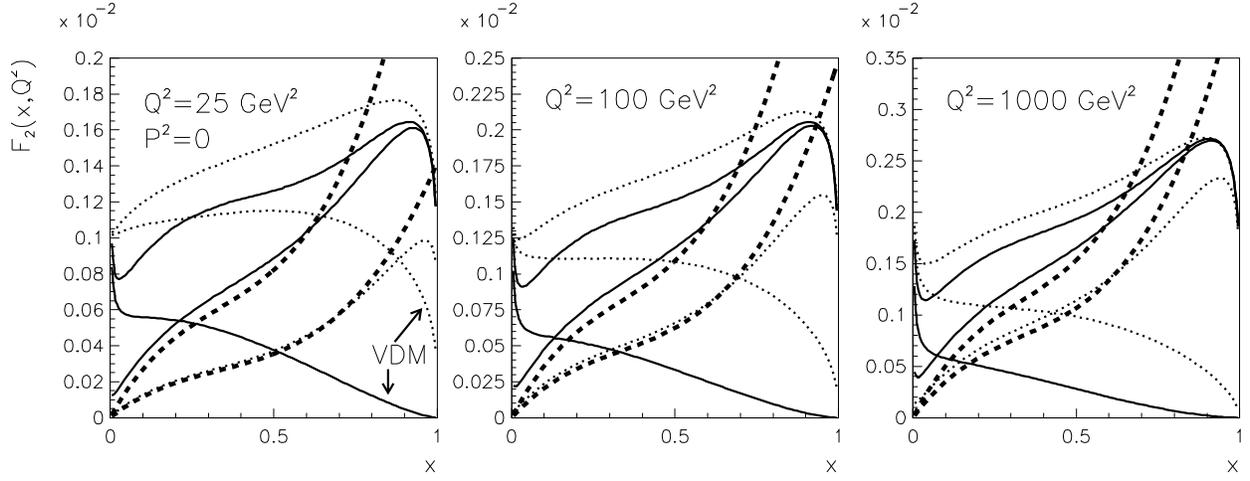,width=\textwidth}
\caption{$F_2^{\gamma}(x,Q^2)$ as given by SaS1D (solid curves) and
SaSD2 (dotted curves) parameterizations. The full results are given
by the upper, the pointlike and VDM contributions parts by two lower
curves. The dashed curves describe the contributions of the
splitting term (\ref{splitterm}).}
\label{f2dreal}
\end{figure}
\begin{figure}\centering
\epsfig{file=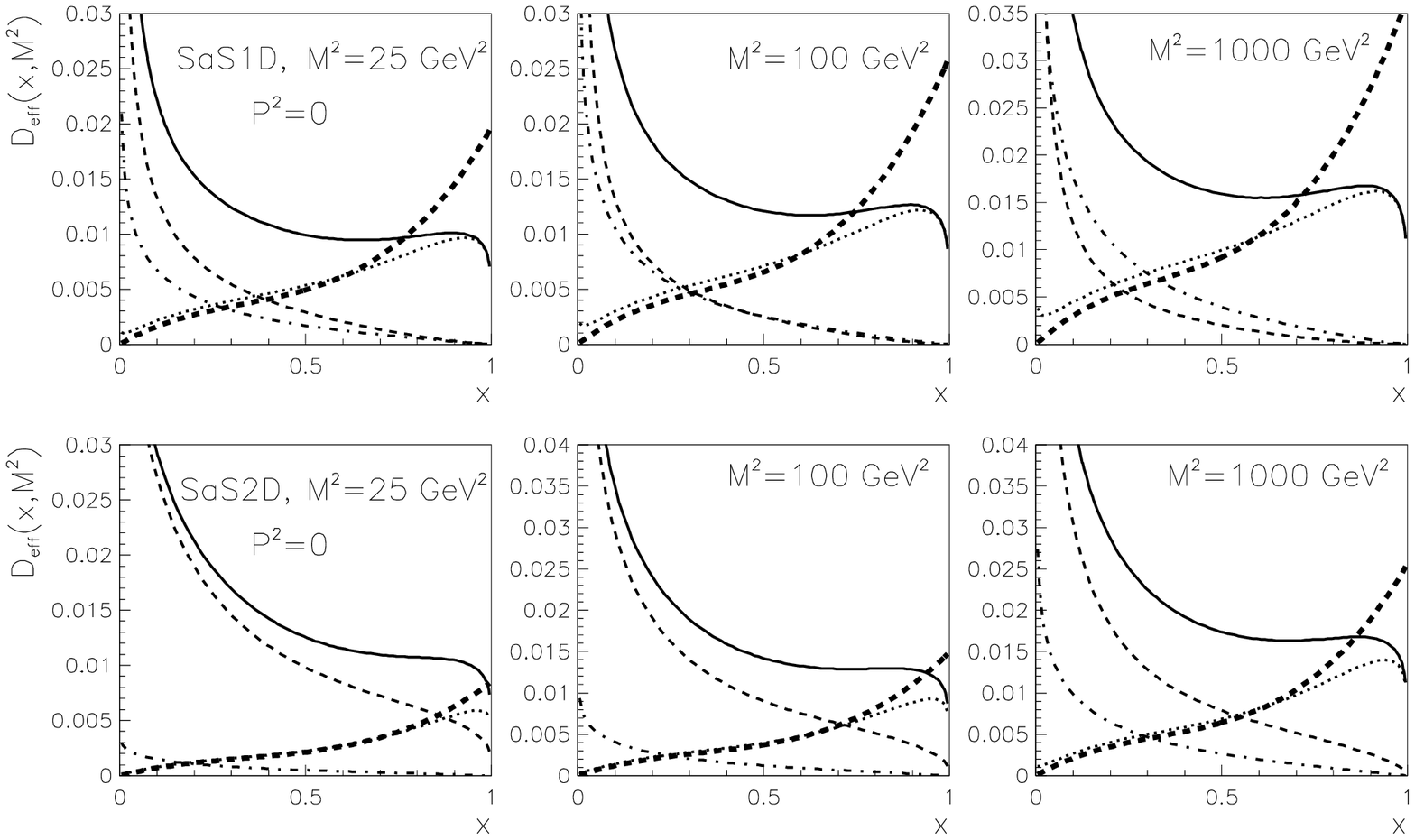,width=\textwidth}
\caption{$D_{\mathrm eff}(x,M^2)$ as given by SaS1D and SaSD2
parameterizations. Solid curves show the full results, dashed ones
the VDM contributions. The pointlike parts are
separated into the contributions of pointlike quarks (dotted curves)
and pointlike gluons (dash--dotted). Thick
dashed curves correspond to the splitting term (\ref{splitterm}).}
\label{eff3dreal}
\end{figure}

In Fig. \ref{sasd1real} distribution functions
$xu(x,M^2)$, $xc(x,M^2)$ and $xG(x,M^2)$ as given by SaS1D and
SaS2D parameterizations for $M^2=100$ GeV$^2$ are compared.
To see how much the
resummation of gluon radiation modifies the first term in
(\ref{resummation}) we also plot the corresponding splitting terms
\begin{equation}
q^{\mathrm split}(x,M_0^2,M^2)\equiv
\frac{\alpha}{2\pi}3e_q^2\left(x^2+(1-x)^2\right)\ln\frac{M^2}{M_0^2}.
\label{splitterm}
\end{equation}
In Fig. \ref{scaledependence} the scale
dependence of VDM and pointlike parts of $u$ and $c$ quark and
gluon distribution functions is displayed.
In the upper six plots we compare
them as a function of $x$ at $M^2=25,100,1000$ GeV$^2$, while in
the lower three plots the same distributions are rescaled by the
factor $\ln(M^2/M_0^2)$. Dividing out this dominant scale
dependence allows us to compare the results directly to curves in
Fig. \ref{anskns}. Finally, in Figs. \ref{f2dreal} and
\ref{eff3dreal} SaS predictions for two physical
quantities, $F_2^{\gamma}$ and effective parton distribution
function $D_{\mathrm {eff}}$, relevant for jet production in
$\gamma$p and $\gamma\gamma$ collisions,
\begin{eqnarray}
F_2^{\gamma}(x,Q^2)& = & \sum_{q}2x e_q^2 q(x,Q^2),
\label{F2PM}\\
D_{\mathrm eff}(x,M^2) & \equiv &
\sum_{i=1}^{n_f}\left(q_i(x,M^2)+\overline{q}_i(x,M^2)\right)+
\frac{9}{4}G(x,M^2)
\label{deff}
\end{eqnarray}
are displayed and compared to results
corresponding to the splitting term (\ref{splitterm}). Figures
\ref{sasd1real}--\ref{eff3dreal} illustrate several important
properties of PDF of the real photon:
\begin{itemize}
\item There is a huge difference between the importance
of the VDM components of light quark and gluon distribution
functions in SaS1D and SaS2D parameterizations: while for SaS2D the
VDM component is dominant up to $x\doteq 0.75$, for SaS1D the
pointlike one takes over above $x\doteq 0.1$!
\item Factorization scale dependence of VDM and
pointlike parts differ substantially. While the former exhibits the
pattern of scaling violations typical for hadrons, the latter
rises, for quarks as well as gluons, with $M$ for all $x$. For
pointlike gluons this holds despite the fact that $G^{\mathrm
{PL}}(x,M^2)$ satisfies at the LO standard homogeneous evolution
equation and is due to the fact that the evolution of $G^{\mathrm
{PL}}(x,M^2)$ is driven by the corresponding increase of
$\Sigma^{\mathrm {PL}}(x,M^2)$.
\item The approach of $q^{\mathrm {PL}}(x,M^2)$ to the asymptotic
pointlike solution (\ref{asymptotic}) is slow
\footnote{
The fact that for the c--quark the SaS2D parameterization is close
to the curve corresponding to $q^{\mathrm {split}}(x,M_0^2,M^2)$,
whereas SaS1D lies substantially below, reflects the fact that
$M_0>m_c$ for SaS2D, but $M_0<m_c$ for SaS1D}. For experimentally
relevant values of
$M$ and except for $x$ close to $1$ the effects of multiple gluon
emission on pointlike quarks are thus small.
\item As the factorization scale $M$ increases the VDM parts of both
quark and gluon distribution functions decrease relative to the
pointlike ones, except in the region of very small $x$.
\item Despite huge differences between SaS1D and SaS2D
parameterizations in the decomposition of quark and gluon
distributions into their VDM and pointlike parts, their
predictions for physical observables  $F_2^{\gamma}$ and
$D_{\mathrm {eff}}$ are much closer.
\item The most prominent effects of multiple parton emission on
physical quantities appear to be the behaviour of
$F_2^{\gamma}(x,Q^2)$ at large $x$ and the contribution of
pointlike gluons to jet cross--sections, approximately described by
$D_{\mathrm {eff}}(x,M^2)$.
\end{itemize}

\section{PDF of the virtual photon}
For the virtual photon the initial state singularity due to the
splitting $\gamma^*\rightarrow q\overline{q}$
is shielded off by the nonzero initial photon virtuality $P^2$ and
therefore in principle
the concept of PDF does not have to be introduced.
In practice this requires, roughly, $P^2>1$ GeV$^2$, where
perturbative QCD becomes applicable.
Nevertheless, PDF still turn out to be phenomenologically very
useful because
\begin{itemize}
\item their pointlike parts include the resummation
of parts of higher order QCD corrections,
\item the hadronic parts, though decreasing rapidly with
$P^2$, are still vital at very small $x$.
\end{itemize}
Both of these aspects define the ``nontrivial'' structure of the
virtual photon in the sense that they are not included in the
splitting term (\ref{splitterm}) and thus are not part of existing
NLO unsubtracted direct photon calculations. One might argue that
the calculable effects of resummation should be considered as
``interaction'' rather than ``structure'', but their uniqueness
makes it natural to describe them in terms of PDF.

\subsection{Equivalent photon approximation}
All the present knowledge of the structure of the photon
comes from experiments at the ep and e$^+$e$^-$ colliders, where
the incoming leptons act as sources of transverse and longitudinal
virtual photons. To order $\alpha$ their respective
unintegrated fluxes are given as
\begin{eqnarray}
f^{\gamma}_{T}(y,P^2) & = & \frac{\alpha}{2\pi}
\left(\frac{1+(1-y)^2)}{y}\frac{1}{P^2}-\frac{2m_{\mathrm e}
^2 y}{P^4}\right),
\label{fluxT} \\
f^{\gamma}_{L}(y,P^2) & = & \frac{\alpha}{2\pi}
\frac{2(1-y)}{y}\frac{1}{P^2}.
\label{fluxL}
\end{eqnarray}
The transverse and longitudinal fluxes thus coincide at $y=0$,
while at $y=1$, $f_{L}^{\gamma}$ vanishes. The $1/P^2$ dependence
of the first terms in (\ref{fluxT}-\ref{fluxL})
results from the fact that in both cases the vertex where photon is
emitted is proportional to $P^2$. This is due to helicity
conservation for the transverse photon and gauge invariance for the
longitudinal one. The term proportional to $m_{\mathrm e}^2/P^4$ in
(\ref{fluxT}) results from the fact that the helicity conservation
at the e$\gamma$e vertex is violated by terms proportional to
electron mass. No such violation is permitted in the case of gauge
invariance, hence the absence of such term in (\ref{fluxL}).
Note that while for $P^2\gg m_{\mathrm e}^2$
the second term in (\ref{fluxT})
is negligible with respect to the leading $1/P^2$ one, close to
$P^2_{\mathrm min}=m_e^2 y^2/(1-y)$ their
ratio is finite and approaches $2(1-y)/(1+(1-y)^2)$.

\subsection{Lessons from QED}
The definition and evaluation of
quark distribution functions of the virtual
photon in pure QED serves as a useful guide to parton model
predictions of virtuality dependence of the pointlike part
of quark distribution functions of the virtual photon.
In pure QED and to order $\alpha$ the probability of
finding inside the photon of virtuality $P^2$ a quark with mass
$m_q$, electric charge $e_q$, momentum fraction $x$ and virtuality
$\tau=m_q^2-k^2\le M^2$ is defined as ($k$ denotes its four-momentum)
\begin{equation}
q_{\mathrm {QED}}(x,m^2_q,P^2,M^2)\equiv
\left(\frac{\alpha}{2\pi}3e_q^2\right)
\int_{\tau^{\mathrm {min}}}^{M^2}
\frac{W(x,m^2_q,P^2)}{\tau^2}{\mathrm d}\tau,
\label{general}
\end{equation}
where the function $W(x,m_q^2,P^2,M^2)$ can in general be written as
\begin{eqnarray}
 W(x,m^2_q,P^2,M^2)& =
&f(x)\frac{p_T^2}{1-x}+g(x)m^2_q+
h(x)P^2+\cdots
\nonumber \\
& = & f(x)\tau+\left(g(x)-\frac{f(x)}{1-x}\right)m^2_q+
\left(h(x)-xf(x)\right)P^2+\cdots.
\label{W}
\end{eqnarray}
In the collinear kinematics, which is relevant for finding the
lower limit on $\tau$, the values of $m_q,~x,~\tau$ and $p_T$ are
related to initial photon virtuality $P^2$ as follows
\begin{equation}
\tau=xP^2+\frac{m_q^2+p_T^2}{1-x},~~~
\Rightarrow ~~~\tau^{\mathrm {min}}=xP^2+\frac{m_q^2}{1-x}.
\label{kinematics}
\end{equation}
The functions $f(x),g(x)$ and $h(x)$, which determine the terms
singular at small $\tau$, are unique functions that have a clear
parton model interpretation: so long as $\tau\ll M^2$ eq.
(\ref{general}) describes the flux of quarks
that are almost collinear with the incoming photon and ``live''
long with respect to $1/M$.
On the other hand, the terms indicated in (\ref{W}) by dots
are of the type $\tau^{k+1}/s^k,k\ge 1$,
which upon insertion into (\ref{general}) yield contributions that
are not singular at $\tau=0$ and therefore do not admit simple
parton model interpretation. In
principle we can include in the definition (\ref{general})
even part of these
{\em nonpartonic} contributions, but we prefer not do that.
Substituting (\ref{W}) into
(\ref{general}) and performing the integration gives, in units of
$3e_q^2\alpha/2\pi$,
\begin{equation}
q_{\mathrm {QED}}(x,m_q^2,P^2,M^2)=
f(x)\ln\left(\frac{M^2}{\tau^{\mathrm {min}}}\right)+
\left[-f(x)+\frac{g(x)m_q^2+h(x)P^2}{\tau^{\mathrm {min}}}
\right]
\left(1-\frac{\tau^{\mathrm {min}}}{M^2}\right).
\label{fullresult}
\end{equation}
In practical applications the factorization scale M is identified
with some kinematical variable characterizing hardness of
the collision, like $\sqrt{Q^2}$ in DIS or
$E_T^{\mathrm jet}$ in jet production.
For $\tau^{\mathrm {min}}\ll
 M^2$ the expression (\ref{fullresult}) simplifies to
\begin{equation}
q_{\mathrm {QED}}(x,m_q^2,P^2,M^2)  =  f(x)\ln\left(
\frac{M^2}{xP^2+m_q^2/(1-x)}\right)-f(x)+
\frac{g(x)m_q^2+h(x)P^2}{xP^2+m_q^2/(1-x)}.
\label{finalresult}
\end{equation}
For $x(1-x)P^2\gg m_q^2$ this expression reduces further to
\begin{equation}
q_{\mathrm {QED}}(x,0,P^2,M^2)=f(x)\ln\left(\frac{M^2}{xP^2}\right)
-f(x)+\frac{h(x)}{x}.
\label{virtualphoton}
\end{equation}
Provided $m_q^2\neq 0$ (\ref{finalresult}) has a finite limit for
$P^2\rightarrow 0$, corresponding to the real photon
\begin{equation}
q_{\mathrm
{QED}}(x,m_q^2,0,M^2)=f(x)\ln\left(\frac{M^2(1-x)}{m_q^2}\right)
-f(x)+g(x)(1-x).
\label{realphoton}
\end{equation}
As in the case of the photon fluxes (\ref{fluxT}-\ref{fluxL})
the logarithmic term, dominant for large $M^2$,
as well as the ``constant'' terms, proportional to $f(x),g(x)$ and
$h(x)$, come entirely from the integration region close to
$\tau^{\mathrm {min}}$ and are therefore unique. At
$\tau=\tau^{\mathrm {min}}$ both types of the singular terms, i.e.
$1/\tau$ or $1/\tau^2$, are of the same order but the faster
fall--off of the $1/\tau^2$ terms implies that for large $M^2$ the
integral over $\tau$, which gives (\ref{fullresult}), is dominated
by the weaker singularity $1/\tau$. In other words, while the
logarithmic term is dominant at large $M^2$, the constant terms
resulting from nonzero $m^2$ and $P^2$ come from the kinematical
configurations which are even more collinear, and thus more
partonic, than those giving the
logarithmic term. The analysis of the vertex $\gamma^*\rightarrow
q\overline{q}$ in collinear kinematics yields \cite{factor}
\begin{equation}
\begin{array}{lll}
 f_T(x)=x^2+(1-x)^2, & g_T(x)=
{\displaystyle
 \frac{\displaystyle 1}{\displaystyle 1-x}
}, & h_T(x)=0, \\[0.4cm]
 f_L(x)=0, & g_L(x)=0, & h_L(x)=4x^2(1-x).
\label{fghTL}
\end{array}
\end{equation}
The expressions (\ref{fullresult}-\ref{finalresult})
exhibit explicitly the smooth transition between quark distribution
functions of the virtual and real photon. This transition is
governed by the ratio $P^2/m^2$, which underlines why in QED
fermion masses are vital. On the other hand, as $P^2$ (or more
precisely $\tau^{\mathrm min}$) increases toward the factorization
scale $M^2$, the above expressions for the quark distribution
functions of the virtual photon vanish. This property holds not
only for the logarithmic term but also for the ``constant''
terms and has a clear intuitive content: virtual photon with
lifetime $1/P\ll 1/M$ does not contain partons living
long enough to take part in the collision characterized by
the interaction time $1/M$.

For virtual photon and $x(1-x)P^2\gg m^2$ the coefficient functions
$C_{\gamma}^T(x,P^2,Q/M),C_{\gamma}^L(x,P^2,Q/M)$ for transverse
and longitudinal target photon polarizations are given as
\cite{russians}
\begin{eqnarray}
C_{\gamma,T}^{(0)}(x,P^2,1)&
= & 3
\left[(x^2+(1-x)^2)\ln\frac{1}{x^2}+8x(1-x)-2\right],
\label{cgammaTvirtual} \\
C_{\gamma,L}^{(0)}(x,P^2,1)& = & 4x(1-x),
\label{cgammaLvirtual}
\end{eqnarray}
whereas for the real photon, i.e. for $P^2=0$
\begin{eqnarray}
C_{\gamma,T}^{(0)}(x,0,1)& = &
3\left[(x^2+(1-x)^2)\ln\frac{1-x}{x}+8x(1-x)-1\right],
\label{cgammaTreal} \\
C_{\gamma,L}^{(0)}(x,0,1)& = & 0
\label{cgammaLreal}
\end{eqnarray}
The origins of the nonlogarithmic parts of $C^{(0)}_{\gamma}$ in
(\ref{cgammaTvirtual}-\ref{cgammaTreal}) can then be identified as
follows:
\begin{eqnarray}
-1+8x(1-x) & = & \underbrace{-2+8x(1-x)}_{\mathrm {for~massless~quark}}+
\underbrace{1}_{g_T(x)(1-x)} \nonumber \\
 & = & \underbrace{-1+6x(1-x)}_{\mathrm {nonpartonic~part}}-
\underbrace{(x^2+(1-x)^2)}_{f_T(x)}+\underbrace{1}_
{g_T(x)(1-x)}~~~~~~~~~
\label{intreal}
\end{eqnarray}
The nonpartonic part itself can be separated into two pieces, coming from
the interaction of the transverse target photon with transverse and
longitudinal probing one
\begin{equation}
-1+6x(1-x)  =  \underbrace{-1+2x(1-x)}_{\mathrm{from~}\sigma_{TT}}
-\underbrace{4x(1-x)}_{\mathrm{from~}\sigma_{LT}}
\label{nonpartonic}
\end{equation}
Except for a brief comment in the next subsection, we shall consider
throughout the rest of this paper the contributions of the transverse
polarization of the target virtual photon only. The importance of
including in analyses of hard collisions of virtual photons the
contributions of $\gamma_L^*$ will be discussed in detail in separate
publication \cite{long}.

\subsection{What is measured in DIS on virtual photons?}
In experiments at e$^+$e$^-$ colliders the
structure of the photon has been investigated via standard DIS on the
photon with small but nonzero virtuality $P^2$. The resulting data
were used in \cite{GRS} to determine PDF of the virtual photon.
In these analyses $C^{(0)}_{\gamma}$ was taken in the form
\begin{equation}
C_{\gamma}^{(0)}(x,P^2,1)=3
\left[(x^2+(1-x)^2)\ln\frac{1}{x^2}+6x(1-x)-2\right],
\label{c0virtwrong}
\end{equation}
which, however, does not correspond to the structure function that
is actually measured in e$^+$e$^-$ collisions, but to the following
combination
\begin{equation}
F_{2,\Sigma}^{\gamma}(x,P^2,Q^2)\equiv F_{2,T}^{\gamma}(x,P^2,Q^2)-
\frac{1}{2}F_{2,L}^{\gamma}(x,P^2,Q^2)
\label{fsigma}
\end{equation}
of structure functions corresponding to transverse and longitudinal
polarizations of the target photon. This combination results after
averaging over the target photon polarizations by means of
contraction with the tensor $-g_{\mu\nu}/2$. The expression
$-2+6x(1-x)$ follows also directly from the definition
(\ref{fsigma}) and considerations of the previous Section:
\begin{equation}
-2+6x(1-x) =  \underbrace{-2+8x(1-x)}_{{\mathrm {from}}~\gamma_T}-
\underbrace{2x(1-x)}_{{\mathrm {from}}~\gamma_L/2}.
\label{intvirt}
\end{equation}
Because the fluxes (\ref{fluxT}--\ref{fluxL}) of transverse and
longitudinal photons are different functions of $y$, any complete
analysis of experimental data in terms of the structure functions
$F_{2,T}^{\gamma}(x,P^2,Q^2)$ and $F_{2,L}^{\gamma}(x,P^2,Q^2)$ at
fixed $x,P^2,Q^2$ requires combining data for different $y$.
This is in principle possible, but
experimentally difficult to accomplish. The situation is simpler at
small $y$, where $f_T^{\gamma}(y,P^2)\doteq
f_L^{\gamma}(y,P^2)=f^{\gamma}(y,P^2)$, and the data therefore
correspond to the convolution of $f^{\gamma}(y,P^2)$ with the sum
$F_{2,T}^{\gamma}+F_{2,L}^{\gamma}$. The nonlogarithmic term in
$C^{(0)}_{\gamma}$ corresponding to this combination is, however, not
$-2+6x(1-x)$, as used in \cite{GRS}, but
$-2+12x(1-x)$, the sum of nonlogarithmic terms corresponding to
transverse and longitudinal photons. Numerically the difference
between these two expressions is quite sizable.

Very recently, the GRS group \cite{GRSch} has changed their approach
to the treatment of the target photon polarizations and argued in
favor of neglecting the contribution of $\gamma_L^*$ and using even
for the virtual photon the same form of $C_{\gamma}^{(0)}$ as for the
real one. We disagree with their arguments, but leave the discussion
of this point to future paper \cite{long}.

\subsection{Virtuality dependent PDF}
\begin{figure}
\epsfig{file=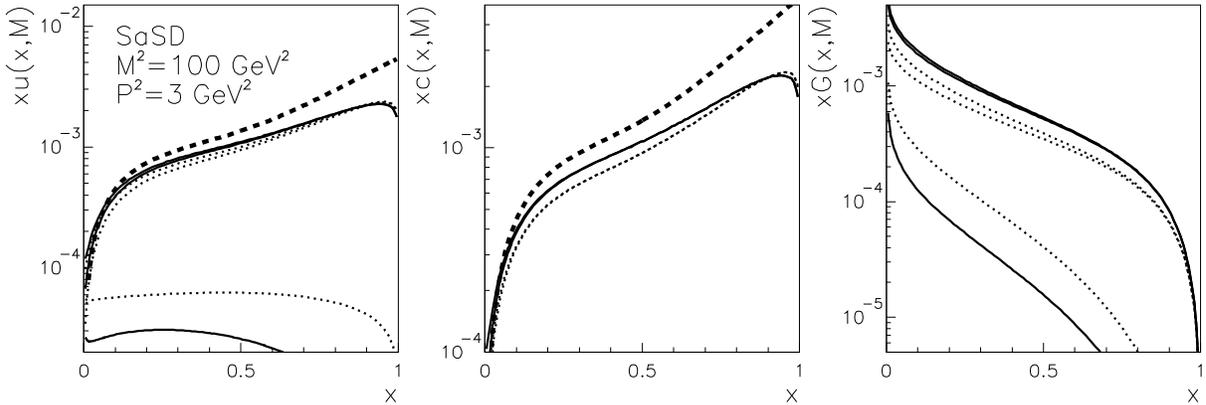,width=\textwidth}
\caption{The same as in Fig. \ref{sasd1real} but for virtual
photon with virtuality $P^2=3$ GeV$^2$. The dashed curves displayed
for $u$ and $c$ quarks correspond to the splitting
term (\ref{splitterm}) with $M_0^2=P^2$.}
\label{sasd1virtual}
\end{figure}

\begin{figure}
\epsfig{file=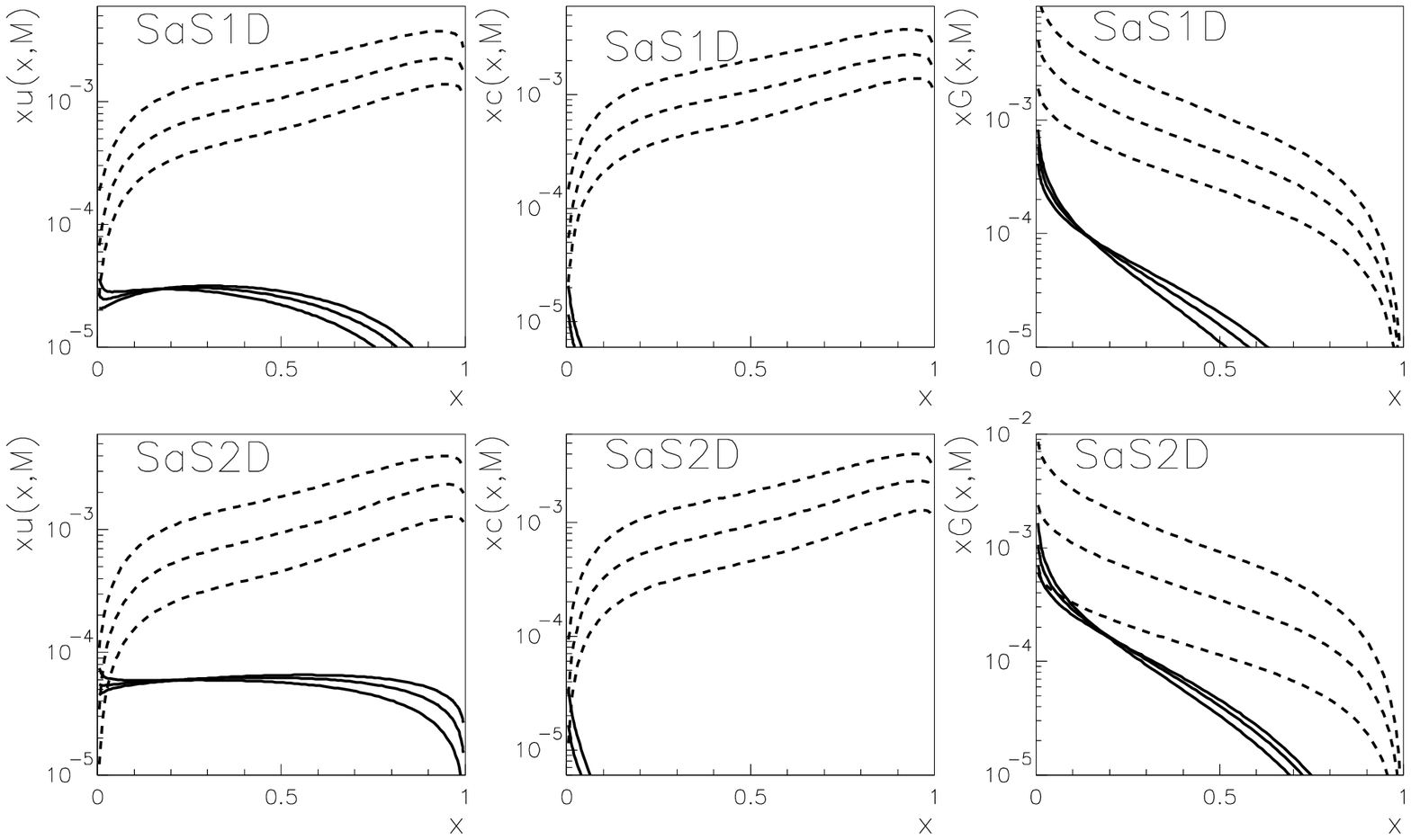,width=\textwidth}
\caption{The same as in Fig. \ref{scaledependence} but for virtual
photon with $P^2=3$ GeV$^2$.}
\label{sasq2p13}
\end{figure}

\begin{figure}\unitlength 1mm
\epsfig{file=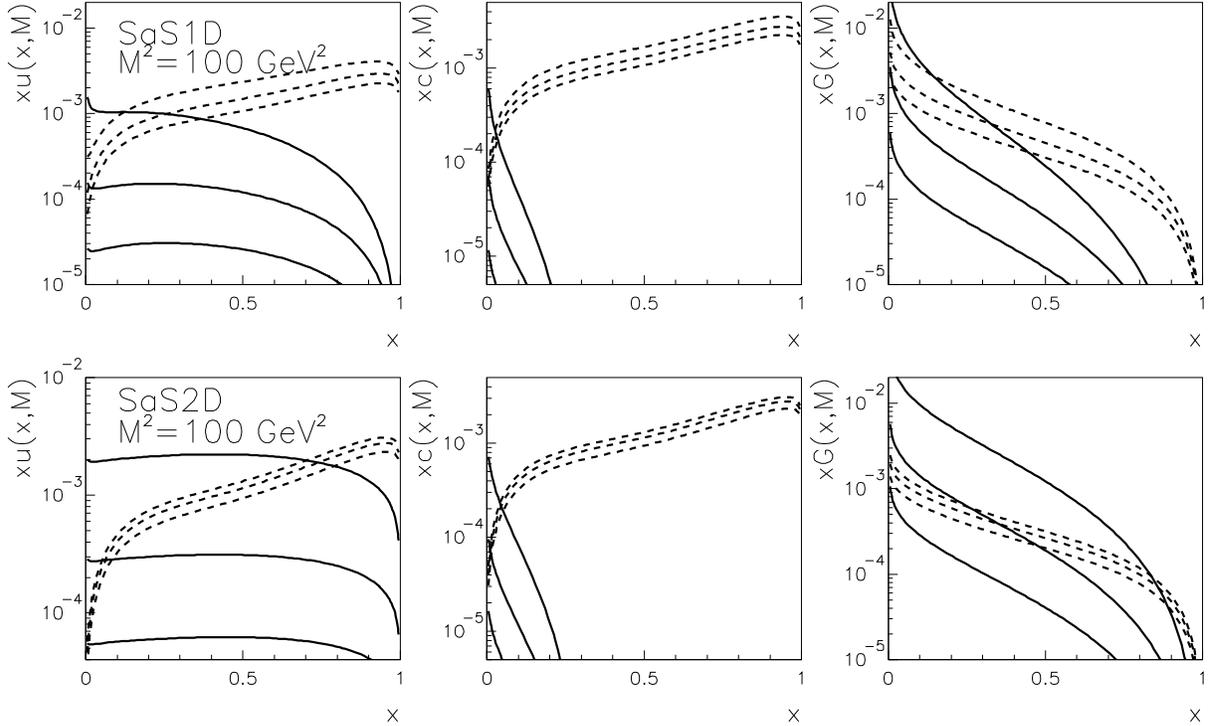,width=\textwidth}
\caption{Virtuality dependence of photonic PDF at $M^2=100$ GeV$^2$.
Dashed curves correspond to the pointlike and solid ones to the VDM
parts, from above for $P^2=0,1,3$ GeV$^2$.}
\label{sasp2}
\end{figure}

\begin{figure}\centering
\epsfig{file=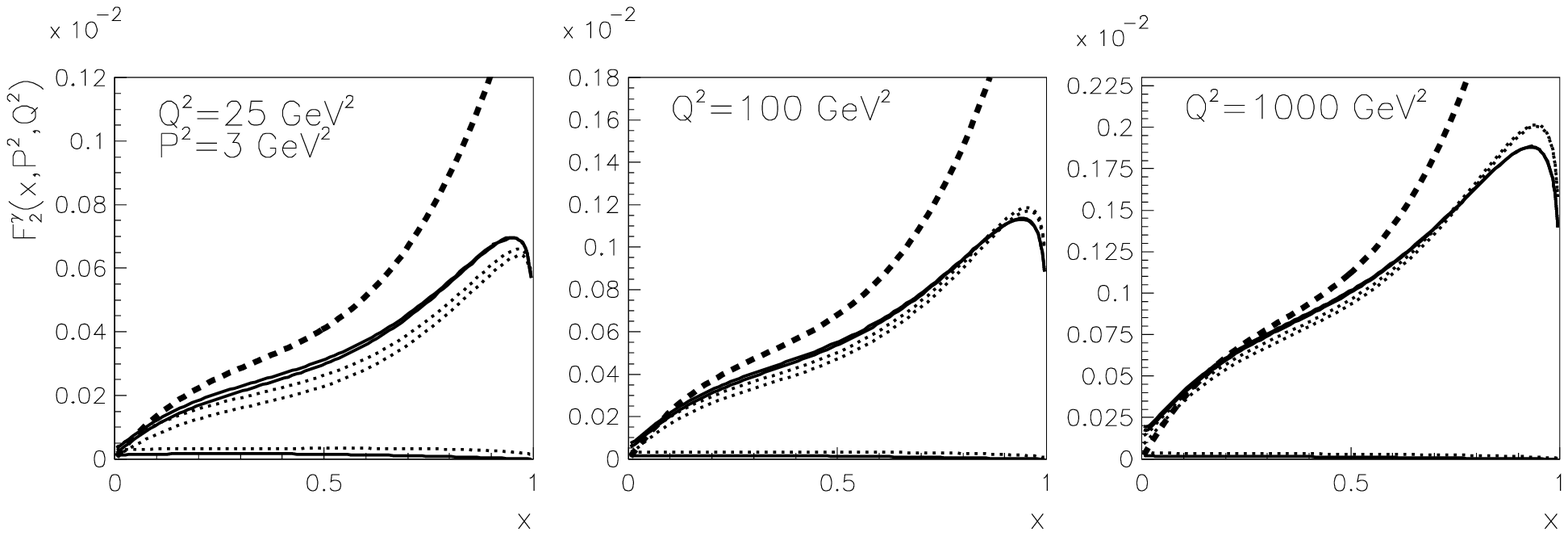,width=\textwidth}
\caption{The same as in Fig. \ref{f2dreal} but for virtual
photon with $P^2=3$ GeV$^2$.}
\label{f2dvirtual}
\end{figure}

\begin{figure}
\epsfig{file=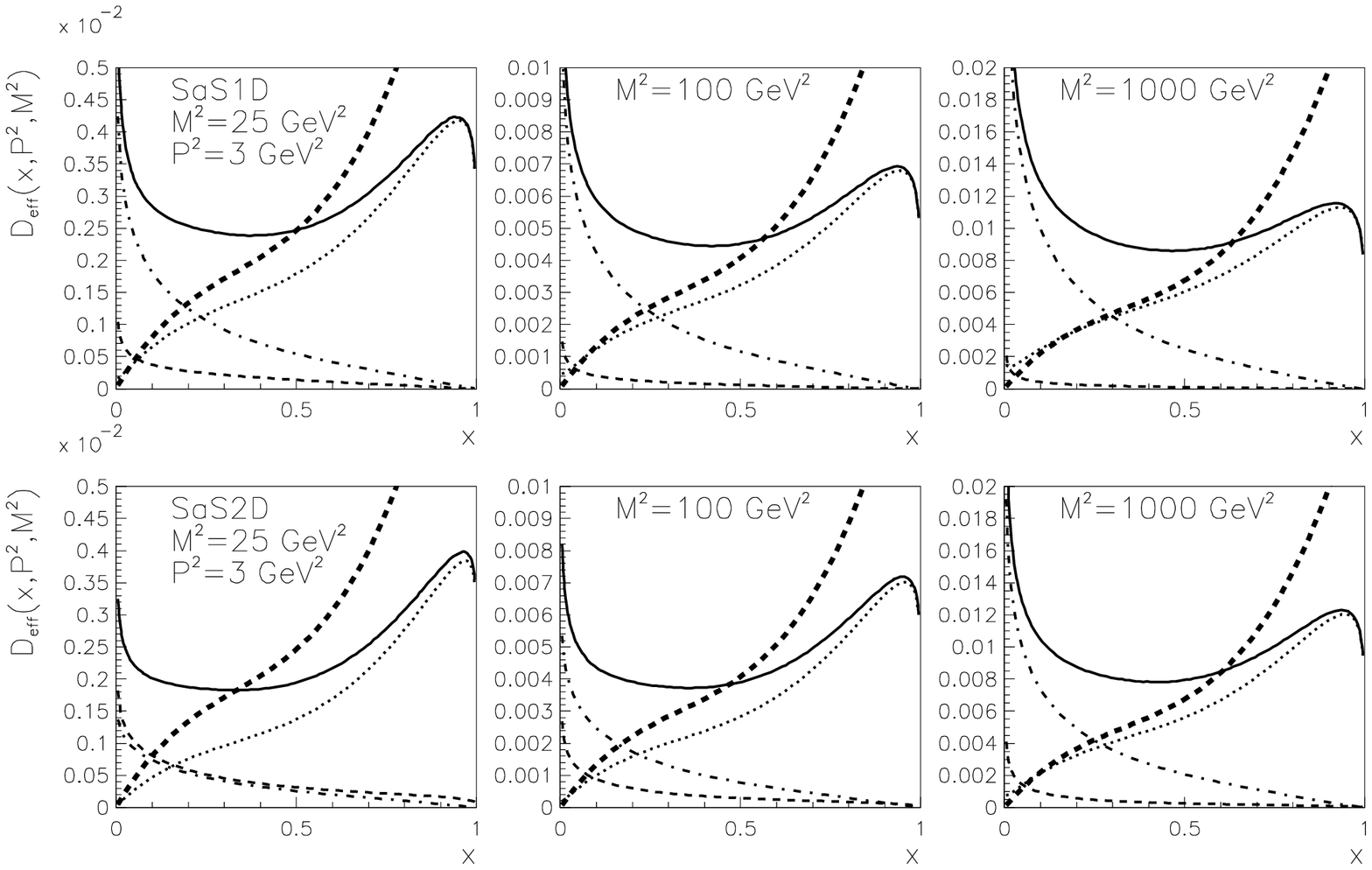,width=\textwidth}
\caption{$D_{\mathrm eff}(x,P^2,M^2)$ as a function of $x$ for
$M^2=25, 100,1000$ GeV$^2$ and $P^2=3$ GeV$^2$. Notation as in
Fig. \ref{eff3dreal}. In the splitting term $M_0^2=P^2$.}
\label{eff3dvirtual}
\end{figure}
In realistic QCD the nonperturbative effects connected with the
confinement, rather than current or constituent
quark masses, are expected to determine
the long--range structure of the photon and hence also the
transition from the virtual photon to the real one. For instance,
within the SaS parameterizations the role of
quark masses is taken over by vector meson masses for the VDM
components and by the initial $M_0$ for the pointlike ones. As in
the case of the real photon, we recall basic features of SaS
parameterizations of the virtual photon, illustrated in Figs.
\ref{sasd1virtual}--\ref{f2dvirtual}.
\begin{itemize}
\item With increasing
$P^2$ the relative importance of VDM parts of both quark and gluon
distribution functions with respect to the corresponding pointlike
ones decreases rapidly. For $M^2\gtrsim 25$ GeV$^2$
the VDM parts of both SaS1D and SaS2D parameterizations become
negligible for $P^2\gtrsim 2$ GeV$^2$, except in the region
of very small $x\lesssim 0.01$. Consequently, also the ambiguity
in the separation (\ref{separation}) is practically negligible
in this region.
\item
The general pattern of scaling violations, illustrated in Fig.
\ref{sasq2p13}, remains the same as for the
real photon, but there is a
subtle difference, best visible when comparing in Fig.
\ref{scaledependence} the rescaled
PDF for $P^2=0$ with those at
$P^2=1,3$ GeV$^2$. While for $P^2=0$, increasing $M^2$ softens the
spectrum towards the asymptotic pointlike form (\ref{asymptotic}),
for $P^2\gtrsim 1$ GeV$^2$ quark distribution functions
increase with $M$ for practically all values of $x$. Moreover,
while for $P^2=0$ the splitting term intersects the SaS1D curves
qualitatively as in Fig. \ref{anskns}, for $P^2\gtrsim 1$ GeV$^2$
it is
above them for all $x$ . These properties reflect the fact that SaS
parameterizations of PDF of the virtual photon do not satisfy the
same evolution equations as PDF of the real one. This difference is
formally of power correction type and thus legitimate within the
leading twist approximation we are working in, but numerically
nonnegligible.
\item As shown in Fig. \ref{sasp2} both VDM and pointlike parts
decrease with increasing $P^2$, but the VDM parts drop much faster
than the pointlike ones.
\end{itemize}
The implications of these properties for physical quantities
$F_2^{\gamma}(x,P^2,Q^2)$ and $D_{\mathrm eff}(x,P^2,M^2)$ are
illustrated by Figs. \ref{eff3dvirtual} and \ref{f2dvirtual}.
For $F_2^{\gamma}(x,P^2,M^2)$ the nontrivial aspects of virtual
photon structure are confined mainly to the region of large $x$,
where they reduce the predictions based on the splitting term
(\ref{splitterm}). Obviously, only the pointlike quarks are
relevant for this effect.
The effects on $D_{\mathrm eff}(x,P^2,M^2)$ worth emphasizing are
the following:
\begin{itemize}
\item For all scales $M$ the contribution of the splitting term is
above the one from pointlike quarks, the gap increasing with
increasing $P^2$ and decreasing $M^2$.
\item Pointlike quarks dominate
$D_{\mathrm eff}(x,P^2,M^2)$ at large $x$, whereas for $x\lesssim
0.3$, most of the pointlike contribution comes from the pointlike
gluons. In particular, the excess of the pointlike contributions to
$D_{\mathrm eff}$ over the contribution of the splitting term,
observed at $x\lesssim 0.5$, comes almost
entirely from the pointlike gluons!
\item For $x\gtrsim 0.6$
the full results are clearly below those given by the splitting
term (\ref{splitterm}) with $M_0^2=P^2$. In this region one
therefore expects the sum of subtracted direct and resolved
contributions to jet cross--sections to be smaller than the results
of unsubtracted direct calculations.
\end{itemize}
So far no restrictions were imposed on transverse energies and
pseudorapidities
\footnote{All quantities correspond to $\gamma^*$p cms.}
of jets produced in $\gamma^*$p collisions.
However, as discussed in more detail in Section 4.3.4, for jet
transverse energies $E_T\gtrsim 5$
GeV, hadronization corrections become intolerably large
and model dependent in the region $\eta\lesssim -2.5$.
On the experimental
side the problems with the reconstruction of jets with
$E_T\gtrsim 5$ GeV in the forward region
restrict the accessible region to $\eta\le 0$. The cuts
imposed on pseudorapidities $\eta^{(i)},i=1,2$ and transverse
energies $E_T^{(i)}$ of two final state partons
\footnote{In realistic QCD analyses of two jets with highest and
second highest $E_T$. Jets with highest and second highest
$E_T$ are labelled ``1'' and ``2''.}
influence strongly the corresponding
distribution in the variable $x_{\gamma}$
\begin{equation}
x_{\gamma}\equiv
\frac{E_T^{(1)}e^{-\eta^{(1)}}+E_T^{(2)}e^{-\eta^{(2)}}}{2E_{\gamma}}
\label{xgammadef}
\end{equation}
used in analyses of dijet data. At the LO and for massless partons
$x_{\gamma}$ defined in (\ref{xgammadef}) coincides with the
conventional fraction $x$ appearing as an argument of photonic PDF.
MC simulations show that for $E_T^{(i)}\ge 5~{\mathrm
{GeV}}$ and $-2.5\le\eta^{(i)}\le 0,i=1,2$,
$\langle x_{\gamma}\rangle\simeq 0.25$, is just the region where
pointlike gluons dominate
$D_{\mathrm {eff}}(x,P^2,M^2)$. This makes jet
production in the region $P^2\gtrsim 1$ a promising place for
identification of nontrivial aspects of PDF of virtual photons.

\section{PDF of the virtual photon in NLO QCD calculations}
Most of the existing information on interactions on virtual photons
comes from the measurements of $F_2^{\gamma}$ at PETRA \cite{PLUTO}
and LEP \cite{L3} and
jet production in ep collisions at HERA \cite{Tania,H1eff}.
In \cite{H1eff}
data on dijet production in the region of virtualities $1\le
P^2\le 80$ GeV$^2$, and for jet transverse energies $E_T^{\mathrm
jet}\ge 5$ GeV have been analyzed within the framework of effective
PDF defined in (\ref{deff}).
This analysis shows that in the kinematical range
$1~{\mathrm {GeV}}^2\lesssim P^2\ll E_T^2$ the data agree
reasonably with the expectations based on SaS parameterizations of
PDF of the virtual photon. The same data may, however, be also
analyzed using the NLO parton level Monte--Carlo programs
\footnote{Three such programs do exist, DISENT, \cite{DISENT},
MEPJET, \cite{MEPJET} and DISASTER \cite{DISASTER}.}
that do not introduce
the concept of PDF of the virtual photon. Nevertheless, so long as
$P^2\ll M^2\approx E_T^2$, the pointlike parts of PDF incorporate
numerically important effects of a part of higher order
corrections, namely those coming from collinear emission of partons
in Fig. \ref{figpl}. This makes the concept of PDF very useful
phenomenologically even for the virtual photon. To illustrate this
point we shall now discuss dijet cross--sections calculated by
means of JETVIP \cite{JETVIP}, currently the only NLO parton
level MC program that includes both the direct and
resolved photon contributions and which can thus be used to
investigate the importance of the latter.

\subsection{Structure of JETVIP}
All the above mentioned parton level NLO MC programs contain the
same full set of partonic cross--sections for the direct photon
contribution up the order $\alpha\alpha_s^2$. Examples of such
diagrams
\footnote{In this subsection the various terms considered will be
characterized by the powers of $\alpha$ and $\alpha_s$ that appear
in hard scattering cross--sections. In the corresponding Feynman
diagrams of Fig. \ref{nnlo} these powers are given by the number of
electromagnetic and strong vertices. Writing
$\sigma(\alpha^j\alpha_s^k)$ will thus mean parton level
cross--sections proportional to $\alpha^j\alpha_s^k$, {\em not}
terms {\em up to} this order! For the latter we shall employ
the standard symbol $O(\alpha_j\alpha_s^k)$. Because PDF of the
photon are proportional to $\alpha$, their convolutions in the
resolved channel with partonic cross--sections $\sigma(\alpha_s^k)$
are of the same order as partonic cross--sections
$\sigma(\alpha\alpha_s^k)$ in the direct channel. For
approximations taking into account the first two or three
powers of $\alpha_s$, in either direct or resolved channel, the
denomination NLO and NNLO will be used.} are in Fig.
\ref{nnlo}a ($\sigma(\alpha\alpha_s)$ tree diagram) and
Fig. \ref{nnlo}b ($\sigma(\alpha\alpha_s^2)$ tree diagram).
Moreover all these programs
contain also one--loop corrections to $\sigma(\alpha\alpha_s)$
tree diagrams. They
differ mainly in the technique used to regularize mass
singularities: MEPJET and JETVIP employ the slicing method whereas
DISENT and DISASTER use the subtraction method. Numerical
comparison of JETVIP and the other codes can be found in
\cite{Duprel,bjorn}. To go one order of
$\alpha_s$ higher and perform complete calculation of the direct
photon contributions up to order $\alpha\alpha_s^3$ would require
evaluating tree diagrams like those in Fig. \ref{nnlo}f,k, as well
as one--loop corrections to diagrams like in Fig. \ref{nnlo}b and
two--loop corrections to diagrams like in Fig. \ref{nnlo}a. So far,
such calculations are not available.

In addition to complete NLO direct photon contribution JETVIP
includes also the resolved photon one. Once the concept of virtual
photon structure is introduced, part of the direct photon
contribution, namely the splitting term (\ref{splitterm}), and in
higher orders also further terms in (\ref{resummation}), is
subtracted from the direct contribution (which for the virtual
photon is nonsingular) and included in PDF appearing in the
resolved photon contribution. To avoid confusion we shall
henceforth use the
term ``direct unsubtracted'' (DIR$_{\mathrm{uns}}$)
to denote NLO direct photon
contributions {\em before} this subtraction and reserve the term
``direct'' for the results {\em after} it. In this
terminology the complete calculations is then given by the sum of
direct and resolved parts and denoted DIR$+$RES.

At the order $\alpha_s^2$ the addition of resolved photon
contribution means including diagrams like those in Fig.
\ref{nnlo}c-e, which involve convolutions of PDF from both proton
and photon sides with $\sigma(\alpha_s^2)$ tree partonic
cross--sections. For a complete $O(\alpha_s^2)$ calculation this is
all that {\em has} to be added to the $O(\alpha\alpha_s^2)$ partonic
cross--sections in direct photon channel.
However, for reasons discussed in detail in the
next subsection, JETVIP includes also NLO resolved contributions,
which involve convolutions of PDF with complete $\sigma(\alpha_s^3)$
partonic cross--sections (examples of relevant diagrams are in
Figs. \ref{nnlo}g--j). This might seem inconsistent as no
corresponding $\sigma(\alpha\alpha_s^3)$ direct photon terms are
included. Nevertheless, this procedure makes sense precisely
because of a clear physical meaning of PDF of the virtual photon!
Numerically, the inclusion of the $\sigma(\alpha_s^3)$
resolved terms turns out to
be very important and in certain parts of the phase space leads to
large increase of JETVIP results compared to those of DISENT,
MEPJET or DISASTER.

\subsection{Factorization mechanism in $\gamma$p interactions}
The main argument for adding $\sigma(\alpha_s^3)$ partonic
cross--sections in the resolved channel to $O(\alpha\alpha_s^2)$
ones in the direct and $O(\alpha_s^2)$ ones in the resolved channels,
is based on specific way factorization mechanism
works for processes involving photons in the initial state.
\begin{figure}\centering
\epsfig{file=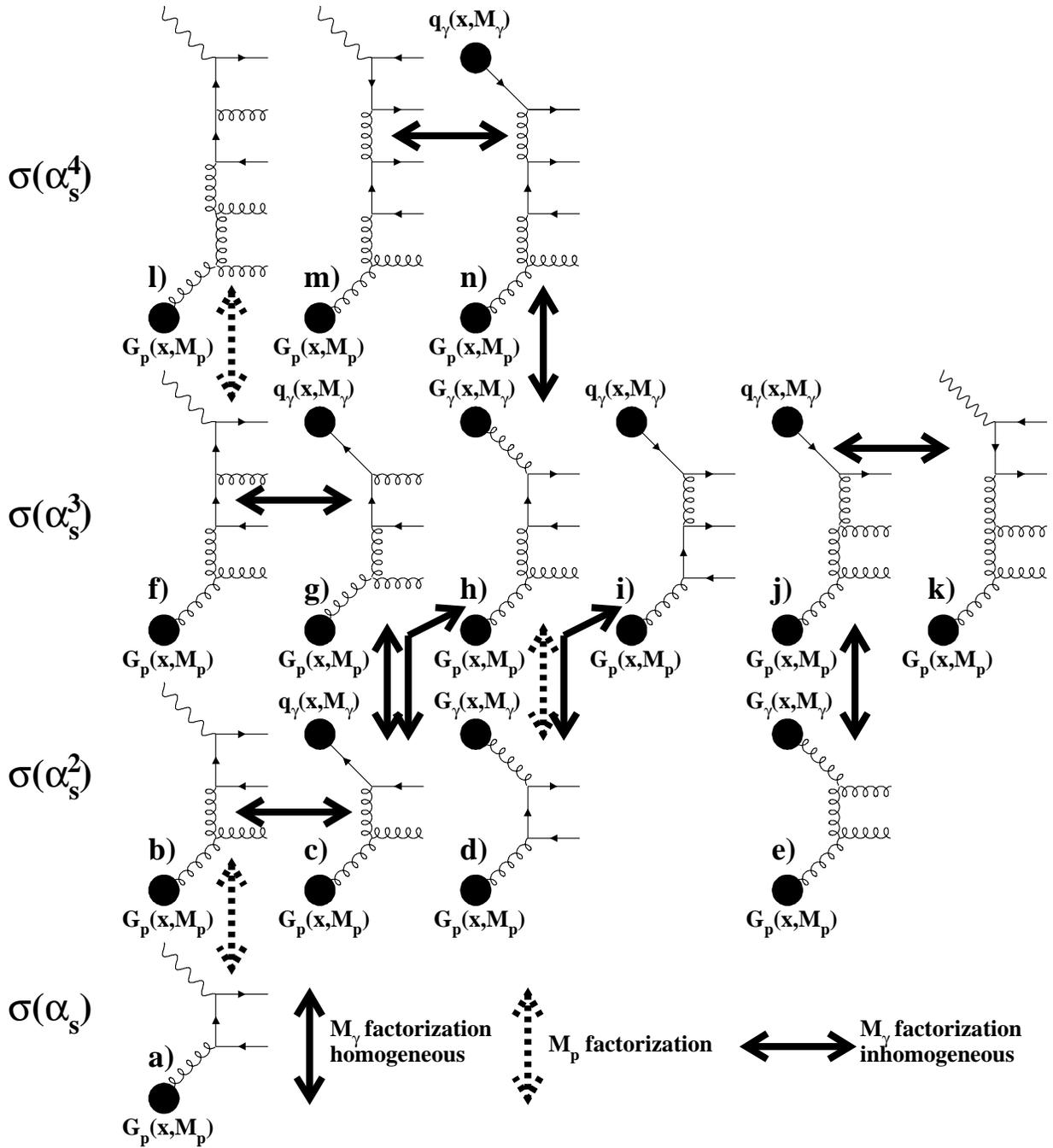,width=\textwidth}
\caption{Examples of diagrams related by factorization mechanism.
Only powers of $\alpha_s$ in parton level cross--sections are counted.}
\label{nnlo}
\end{figure}
First, however, let us recall how the factorization works in hadronic
collisions. Jet cross--sections start as convolutions of
$\sigma(\alpha_s^2)$ partonic cross--sections with PDF of beam and
target hadrons. The factorization scale dependence of these
PDF is cancelled by explicit factorization scale dependence of
higher order partonic cross--sections. This cancellation is exact
provided all orders of perturbation theory
\footnote{In perturbation expansions of partonic cross--sections
as well as splitting functions.} are taken into account,
but only partial in any finite
order approximation, like the NLO one used in analyses of jet
production at FERMILAB. In hadronic collisions the inclusion of
$\sigma(\alpha_s^3)$ partonic cross--sections is thus vital for
compensation of the factorization scale dependence of PDF in
convolutions with $\sigma(\alpha_s^2)$ partonic cross--sections. The
residual factorization scale dependence of such NLO approximations
is formally of the order $\alpha_s^4$ and thus one order of
$\alpha_s$ higher than the terms included in the NLO approximation.

In $\gamma$p collisions (whether of real or virtual photon) the
situation is different due to the presence of inhomogeneous terms
$k_q,k_G$ in the evolution equations
(\ref{Sigmaevolution}--\ref{NSevolution}).
For quark distribution functions (nonsinglet as well as singlet)
the leading term $(\alpha/2\pi)k_q^{(0)}$ in the expansion of
$k_q$ is independent of $\alpha_s$
and consequently part of the factorization
scale dependence of the $\sigma(\alpha_s^2)$ resolved photon
contribution is compensated at the same order $\alpha_s^2$. In
further discussions we shall distinguish two factorization scales:
one ($M_{\gamma}$) for the photon, and the other ($M_p$) for the
proton. The content of the evolution equations
(\ref{Sigmaevolution}--\ref{NSevolution}) is represented
graphically in Fig. \ref{nnlo}, which shows examples of diagrams,
up to order $\alpha_s^4$, relevant for our discussion. Some of
these diagrams are
connected by solid or dashed arrows, representing graphically
the effects of variation of factorization scales $M_{\gamma}$ and
$M_p$ respectively. The vertical dashed arrows connect diagrams
(with lower blobs representing PDF of the proton, denoted
$D_p(x,M_p)$) that differ in partonic cross--sections by one order
of $\alpha_s$, reflecting the fact that the terms on the r.h.s. of
evolution equation for $D_p(x,M_p)$ start at the order $\alpha_s$.
For instance, the $\sigma(\alpha\alpha_s)$ direct photon diagram in
Fig.
\ref{nnlo}a is related by what we call $M_p$--factorization with
$\sigma(\alpha\alpha_s^2)$ direct photon diagram in Fig. \ref{nnlo}b.
Similarly, the $\sigma(\alpha_s^2)$ resolved photon diagram in Fig.
\ref{nnlo}d is related by $M_p$--factorization to $\sigma(\alpha_s^3)$
resolved photon diagram in Fig. \ref{nnlo}h. In fact each diagram
of the order $\alpha_s^k$ is related by $M_p$--factorization to two
types of diagrams at order $\alpha_s^{k+1}$, one with quark and the
other with gluon coming from the proton blob. Note that
$M_p$--factorization operates within either direct or resolved
contributions separately, never relating one type of terms with
the other.

For $M_{\gamma}$--factorization this cancellation mechanism has a
new feature. Similarly as for hadrons, the $\sigma(\alpha_s^2)$
resolved photon
diagram in Fig. \ref{nnlo}c is related by what we call {\em
homogeneous} $M_{\gamma}$--factorization to the $\sigma(\alpha_s^3)$
resolved photon diagrams in Fig. \ref{nnlo}g,h, and similar
relation holds also between the diagram in Figs. \ref{nnlo}e and
\ref{nnlo}j. However, the inhomogeneous term in evolution equation
for PDF of the photon implies additional relation (which we call
{\em inhomogeneous} $M_{\gamma}$--factorization) between direct and
resolved photon diagrams at the {\em same} order of $\alpha_s$,
represented by horizontal solid arrows. For instance, the LO
resolved contribution coming from diagram in Fig. \ref{nnlo}c is
related not only by homogeneous $M_{\gamma}$--factorization to the
$\sigma(\alpha_s^3)$ resolved photon diagrams \ref{nnlo}g-h, but also by
inhomogeneous $M_{\gamma}$--factorization to the
$\sigma(\alpha\alpha_s^2)$ direct photon diagram in Fig. \ref{nnlo}b.
Similarly, the $\sigma(\alpha_s^3)$ resolved photon diagram \ref{nnlo}g
is related by homogeneous $M_{\gamma}$--factorization to
$\sigma(\alpha_s^4)$ resolved terms (not shown) and by inhomogeneous
$M_{\gamma}$--factorization to the $\sigma(\alpha\alpha_s^3)$ direct
diagram in Fig. \ref{nnlo}f.

The argument for adding the $\sigma(\alpha_s^3)$ resolved photon
diagrams to the $\sigma(\alpha_s^2)$ resolved photon ones relies
on the fact
that within the resulting set of NLO resolved contributions the
homogeneous $M_{\gamma}$--factorization operates in the same
way as the
$M_p$--factorization does within the set of NLO direct ones!
However, contrary to the hadronic case, the $\sigma(\alpha_s^3)$
 parton level cross--sections do not constitute
after convolutions with photonic PDF a complete set of
$\alpha\alpha_s^3$ contributions, the rest coming from the
$\sigma(\alpha\alpha_s^3)$ direct ones.
As these $\sigma(\alpha\alpha_s^3)$
direct photon contributions have not yet been calculated, the set
of contributions included in JETVIP does not constitute the
complete $O(\alpha\alpha_s^3)$ calculation.
The lacking terms, like that
coming from the diagram \ref{nnlo}f, would provide
cancellation mechanism at the order $\alpha\alpha_s^3$
with respect to the inhomogeneous $M_{\gamma}$--factorization. In
the absence of $\sigma(\alpha\alpha_s^3)$ direct photon calculations,
we thus have two options:
\begin{itemize}
\item To stay within the framework of complete $O(\alpha_s^2)$
calculations, including the LO resolved and NLO direct
contributions, but with no mechanism for the cancellation of
the dependence of PDF of the virtual photon on the factorization
scale $M_{\gamma}$.
\item To add to the previous framework the $\sigma(\alpha_s^3)$
resolved
photon contribution, which provide the necessary cancellation
mechanism with respect to homogeneous $M_{\gamma}$--factorization,
but do not represent a complete set of $\sigma(\alpha_s^3)$
contributions.
\end{itemize}
In our view the second strategy, adopted in JETVIP, is more
appropriate. In fact one can look at $\sigma(\alpha_s^3)$ resolved
photon terms as results of approximate evaluation of the so far
uncalculated $\sigma(\alpha\alpha_s^3)$ direct photon diagrams in the
collinear kinematics. For instance, so long as $P^2\ll
M_{\gamma}^2$ taking into account only the pointlike part of
$D_{\gamma}(x,P^2,M_{\gamma}^2)$ in the upper blob of
Fig. \ref{nnlo}g
should be a good approximation of the contribution of the direct
photon diagram in Fig. \ref{nnlo}f. There are of course
$\sigma(\alpha\alpha_s^3)$ direct photon diagrams that cannot be
approximated in this way, but we are convinced that it
makes sense to build phenomenology on this framework.

For the $\sigma(\alpha_s^2)$ resolved terms the so far unknown
$\sigma(\alpha\alpha_s^3)$ direct photon contributions provide the
first
chance to generate pointlike gluons inside the photon: the gluon in
the upper blob in resolved photon diagram in Fig. \ref{nnlo}e must
be radiated by a quark that comes from primary $\gamma\rightarrow
q\overline{q}$ splitting, for instance as shown in Fig.
\ref{nnlo}j. Note that to get the gluon in $\sigma(\alpha_s^3)$
resolved
photon contributions, for instance in Fig. \ref{nnlo}h, within
collinear limit of direct photon contributions would require
evaluating diagrams, like that in Fig. \ref{nnlo}m, at the order
$\sigma(\alpha\alpha_s^4)$! In summary, although the pointlike parts of
quark and gluon distribution functions of the virtual photon are in
a sense included in higher order perturbative corrections and can
therefore be considered as expressions of ``interactions'' rather
than ``structure'' of the virtual photon, their uniqueness and
phenomenological usefulness definitely warrant their introduction
as well as name.

\subsection{Theoretical uncertainties}
NLO calculations of jet cross--sections are affected by a number
of ambiguities
caused by the truncation of perturbation expansions as well as
by uncertainties related to the input quantities and conversion
of parton level quantities to hadron level observables.

\subsubsection{Choice of PDF}
We have taken CTEQ4M and SAS1D sets of
PDF of the proton and photon respectively as our principal choice.
Both of these sets treat quarks, including the heavy ones, as
massless above their respective mass thresholds, as required by
JETVIP, which uses LO and NLO matrix elements of massless partons.
PDF of the proton are fairly well determined from global analyses
of CTEQ and MRS groups and we have therefore estimated the residual
uncertainty related to the choice of PDF of the proton by comparing
the CTEQ4M results to those obtained with MRS(2R) set. The
differences are very small, between 1\% at $\eta=-2.5$ and 3.5\%
at $\eta=0$, independently of $P^2$.

For the photon we took the Schuler-Sj\"{o}strand parameterizations
for two reasons.
First, they provide separate parameterizations of VDM and pointlike
components of all PDF, which is crucial for physically transparent
interpretation of JETVIP results. Secondly, they represent the only
set of photonic PDF with physically well motivated virtuality
dependence, which is compatible with the way JETVIP treats heavy
quarks. The GRS sets \cite{GRS,GRSch},
the only other parameterizations with built-in virtuality dependence,
are incompatible with JETVIP because they treat $c$ and $b$ quarks as
massive and, consequently, require calculating their contribution to
physical quantities via the boson gluon or gluon gluon fusion
involving exact massive matrix elements.

\subsubsection{The choice of $n_f$}
JETVIP, as well as other NLO parton calculations of jet
cross--sections, works with a fixed number $n_f$ of massless
quarks, that must be chosen accordingly. This is not a simple
task, as the number of quarks that can be considered effectively
massless depends on kinematical variables characterizing the
hardness of the collision.
Consequently, the optimal choice of $n_f$ may not be unique for the
whole kinematical region under consideration. The
usual procedure is to run such programs for two (or more)
relevant values of $n_f$ and use the ensuing difference as an
estimate of theoretical uncertainty  related to the approximate
treatment of heavy quark contributions.

The number $n_f$ enters NLO calculations in three places:
implicitly in $\alpha_s(\mu)$ and PDF and explicitly in LO and NLO
parton level cross--sections. In our selected region of phase
space the appropriate value of $n_f$ lies somewhere between
$n_f=4$ and $n_f=5$, with the latter value representing the upper
bound on the results (so far unavailable) that would take the
$b$-quark mass effects properly into account. We have therefore
run JETVIP for both $n_f=4$ and $n_f=5$ and compared their
results. They differ, not suprisingly, very little
\footnote{As the contribution of the $b$-quark in both direct and
resolved channels is proportional to its charge, we except it to
amount to about $e_b^2/(e_u^2+e_c^2+e_d^2+e_s^2)=0.1$ of the sum
of light quark ones.}. Explicit calculations give at most $10$\%
in the direct channel and $5$\% in the resolve done. All the
results presented below correspond to $n_f=5$.

\subsubsection{Factorization and renormalization scale dependence}
As mentioned in the previous subsection, proton and photon are
associated in principle with different factorization scales $M_p$
and $M_{\gamma}$, but we followed the standard practice of assuming
$M\equiv M_p=M_{\gamma}$ and set $M=\kappa E_T^{(1)}$.
The factorization scale dependence was quantified by performing
the calculations for $\kappa=0.5,1$ and $\kappa=2$.

The dependence of finite order perturbative calculations on the
renormalization scale $\mu$ is in principle a separate ambiguity,
but we have again followed the common practice of identifying these
two scales $\mu=M$. To reflect this identification, we shall in the
following use the term ``scale dependence'' to describe the dependence
on this common scale.

\subsubsection{Hadronization corrections}
JETVIP, as a well as other NLO codes evaluate jet cross--sections
at the parton level. For a meaningful comparison with experimental
data they must therefore be corrected
for effects describing the conversion of partons to observable
hadrons. These so called hadronization corrections are not
simple to define, but adopting the definition used by
experimentalists \cite{Wobisch} we have found that they
depended sensitively and in a correlated manner on transverse
energies and pseudorapidities of jets. In order to avoid regions
of phase space where they become large we
imposed on both jets the condition $-2.5\le\eta^{(i)}\le 0$.
In this region hadronization corrections are flat in $\eta$ and
do not exceed $10$\%, whereas for $\eta\doteq -2.5$ they
steeply rise with decreasing $\eta$. This by itself would not
require excluding this region, the problem is that in this region
hadronization corrections become also very much model dependent
and therefore impossible to estimate reliably.
Detailed analysis of various aspects of estimating these
hadronization , with particular emphasize on their implication for
jet production at HERA, is contained in \cite{phd}.

\subsubsection{Limitations of JETVIP calculations}
Despite its undisputable advantage over the calculations that do
not introduce the concept of virtual photon structure, also JETVIP
has a drawback because it does not represent a complete NLO QCD
calculation of jets cross--sections. This is true in the
conventional approach to photonic interactions, and even more in
the reformulation suggested by one of us in \cite{jfactor}. In the
conventional approach the incompletness is related to the fact
that there is no NLO parameterization of PDF of the virtual photon
compatible with JETVIP treatment of heavy quarks. Note that in the
standard approach the inclusion of
$\sigma(\alpha_s^3)$ partonic cross--sections in
the resolved photon channel is justified by the claim that their
convolution with photonic PDF are of the same order
$\alpha\alpha_s^2$ as the direct photon ones. In the reformulation
\cite{jfactor} this incompletness has deeper causes. It reflects
the lack of appropriate input PDF but also the fact that a complete NLO
approximation requires the inclusion of direct photon contribution
of the order $\alpha\alpha_s^3$, which is so far not available.
Nevertheless, we reiterate that it makes sense to build phenomenology
upon the current JETVIP framework and the concept of PDF of the
virtual photon is just the necessary tool for accomplishing it.

\subsection{Dijet production at HERA}
We shall now discuss the main features of dijet cross--sections
calculated by means of JETVIP. To make our conclusions potentially
relevant for ongoing analyses of HERA data we have chosen the
following kinematical region
$$E_T^{(1)}\ge E_T^c+\Delta,~E_T^{(2)}\ge E_T^c,
~~~E_T^c=5~{\mathrm{GeV}},
~~\Delta=2~{\mathrm{GeV}}$$
 $$-2.5 \le \eta^{(i)}\le 0,~i=1,2$$
in four windows of photon virtuality
$$1.4\le P^2\le 2.4~{\mathrm {GeV}}^2;
~2.4\le P^2\le 4.4~{\mathrm {GeV}}^2;~
4.4\le P^2\le 10~{\mathrm GeV}^2;
~10\le P^2\le 25~{\mathrm {GeV}}^2$$
The cuts on $E_T$ were chosen in such a way that throughout the region
$P^2\ll E^2_T$, thereby ensuring
that the virtual photon lives long enough for its
``structure'' to develop before the hard scattering takes place.
The asymmetric cut option is appropriate for our decision to plot
the sums of $E_T$ and $\eta$ distributions of the jets with highest
and second highest $E_T$.
The choice of $\Delta=2$ GeV, based on a detailed investigation
\cite{phd} of the dependence of the integral
over the selected region on $\Delta$, avoids the region where
this dependence possesses unphysical features.

In our analysis jets are defined by means of the cone
algorithm. At NLO parton level all jet algorithms are essentially
equivalent to the cone one, supplemented with the parameter
$R_{\mathrm sep}$, introduced in \cite{rsep} in order to bridge the
gap between the application of the cone algorithm to NLO parton
level calculations and to hadronic systems (from data or MC), where
one encounters ambiguities related to seed selection and jet
merging. In a general cone algorithm two objects (partons, hadrons
or calorimetric cells) belong to a jet if they are within the
distance $R$ from the jet center. Their relative distance
satisfies, however, a weaker condition
\begin{equation}
\Delta R_{ij}=\sqrt{(\Delta \eta_{ij})^2+(\Delta \phi_{ij})^2}
\le \frac{E_{T_i}+E_{T_j}}{{\mathrm max}(E_{T_i},E_{T_j})}R.
\end{equation}
The parameter $R_{\mathrm sep}$ governs the maximal distance
between two partons within a single jet, i.e. two partons form a
jet only if their relative distance $\Delta R_{ij}$ satisfies
the condition
\begin{equation}
\Delta R_{ij}\le {\mathrm min}\left[
\frac{E_{T_i}+E_{T_j}}{{\mathrm max}(E_{T_i},E_{T_j})}R,
R_{\mathrm sep}\right].
\end{equation}
The question which value of $R_{\mathrm sep}$ to choose for the
comparison of NLO parton level calculations with the results of the
cone algorithm at the hadron level is nontrivial and we shall
therefore present NLO results for both extreme choices $R_{\mathrm
sep}=R$ and $R_{\mathrm {sep}}=2R$. To define momenta of jets
JETVIP uses the standard $E_T$--weighting recombination procedure,
which leads to massless jets.

\subsection{Results}
To asses phenomenological importance of the concept
of PDF of virtual photons we now compare JETVIP results
obtained in the DIR$_{\mathrm{uns}}$ mode, where this concept is
not introduced at all, with those of the DIR$+$RES one, in which the
contribution of the resolved photon is added to the subtracted
direct one. The difference between these two results measures
the nontrivial aspects of PDF of the virtual photon.

\begin{figure}\unitlength 1mm
\begin{picture}(160,80)
\put(0,0){\epsfig{file=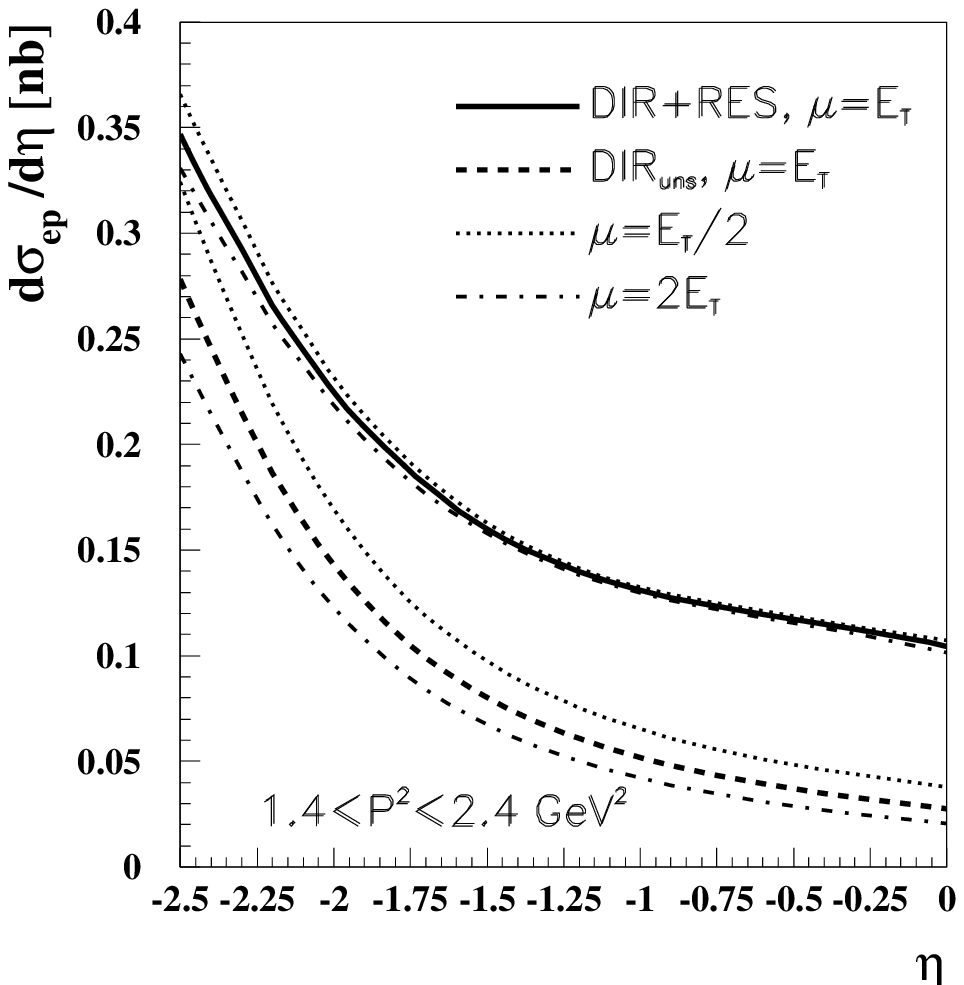,width=7.5cm}}
\put(80,0){\epsfig{file=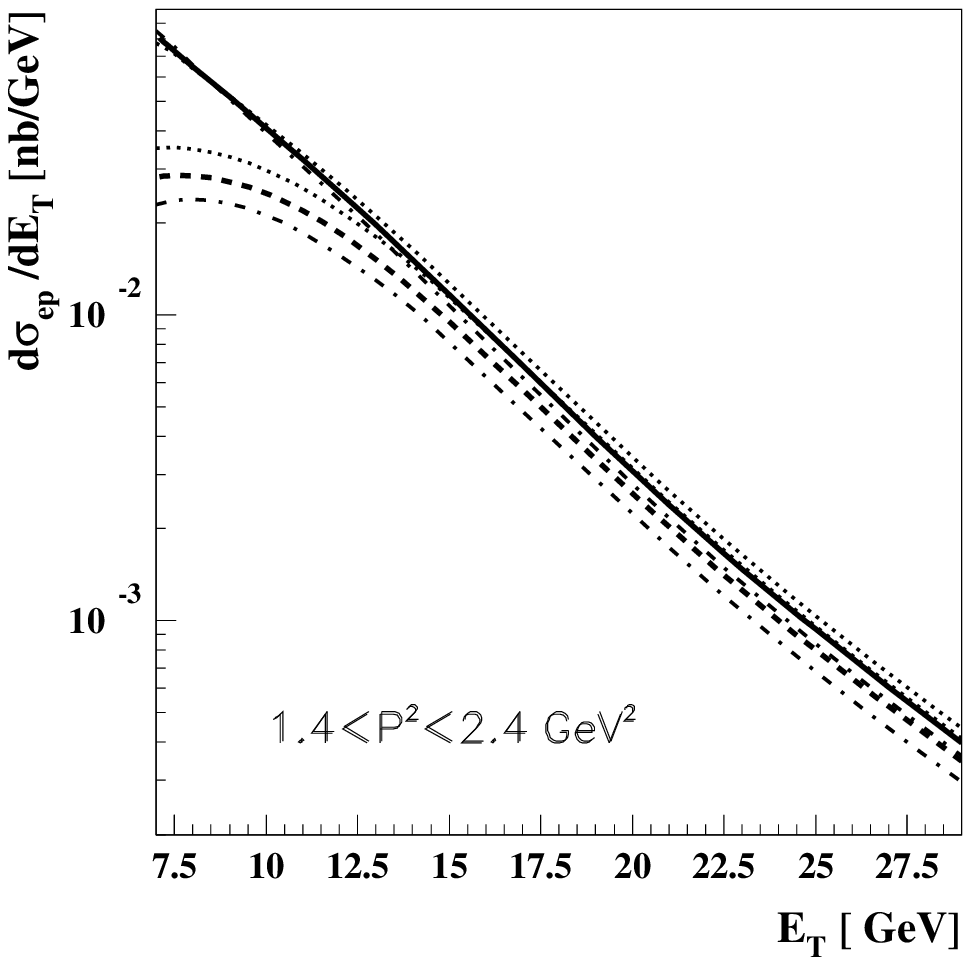,width=7.5 cm}}
\end{picture}
\caption{Scale dependence of the distributions
${\mathrm d}\sigma/{\mathrm d}\eta$ and ${\mathrm d}\sigma/{\mathrm
d}E_T$ at the NLO. All curves correspond to $R_{\mathrm {sep}}=2R$.}
\label{win1}
\end{figure}
\begin{figure}
\epsfig{file=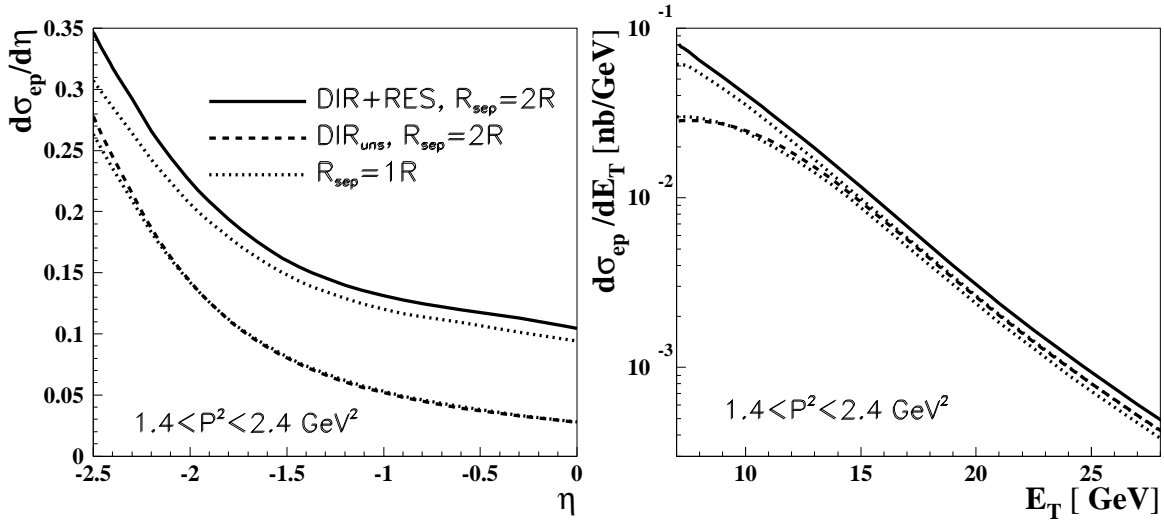,width=\textwidth}
\caption{$R_{\mathrm sep}$ dependence of
${\mathrm d}\sigma/{\mathrm d}\eta$ and ${\mathrm d}\sigma/{\mathrm
d}E_T$ distributions. All curves correspond to $\mu=E_T$.}
\label{drsep}
\end{figure}

\begin{figure}
\epsfig{file=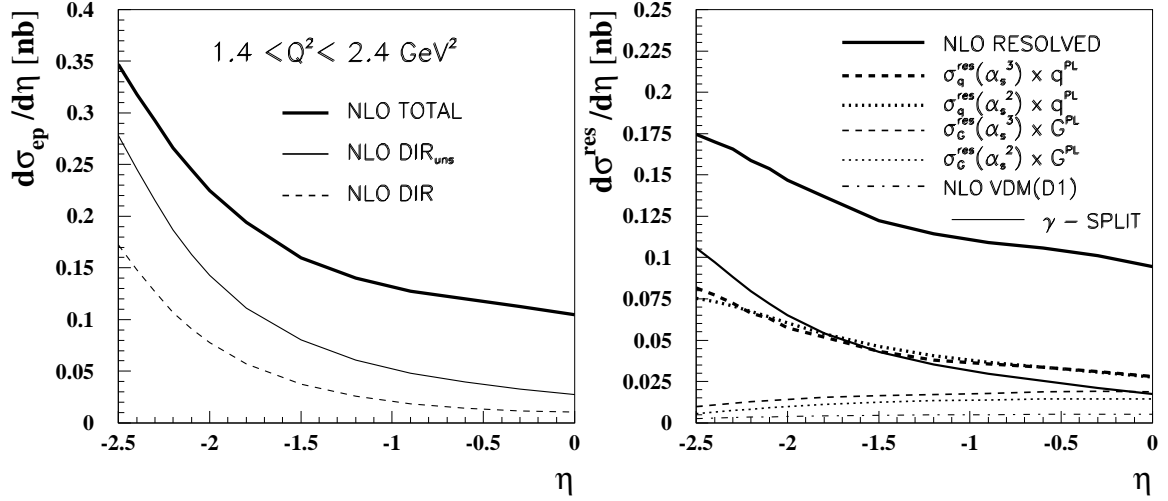,width=\textwidth}
\caption{Comparison of DIR$+$RES, DIR$_{\mathrm uns}$ and
DIR results for ${\mathrm d}\sigma/{\mathrm d}\eta$
(left plots) and individual contributions to
${\mathrm d}\sigma^{\mathrm {res}}/{\mathrm d}\eta$,
described in the text (right plots). The thin solid curve corresponds
to convolution of the splitting term (\ref{splitterm}) with
$\sigma^{\mathrm{res}}_q(\alpha_s^2)$ parton level cross--sections.}
\label{reseta1}
\end{figure}

\begin{figure}\unitlength 1mm
\begin{picture}(160,60)
\put(0,0){\epsfig{file=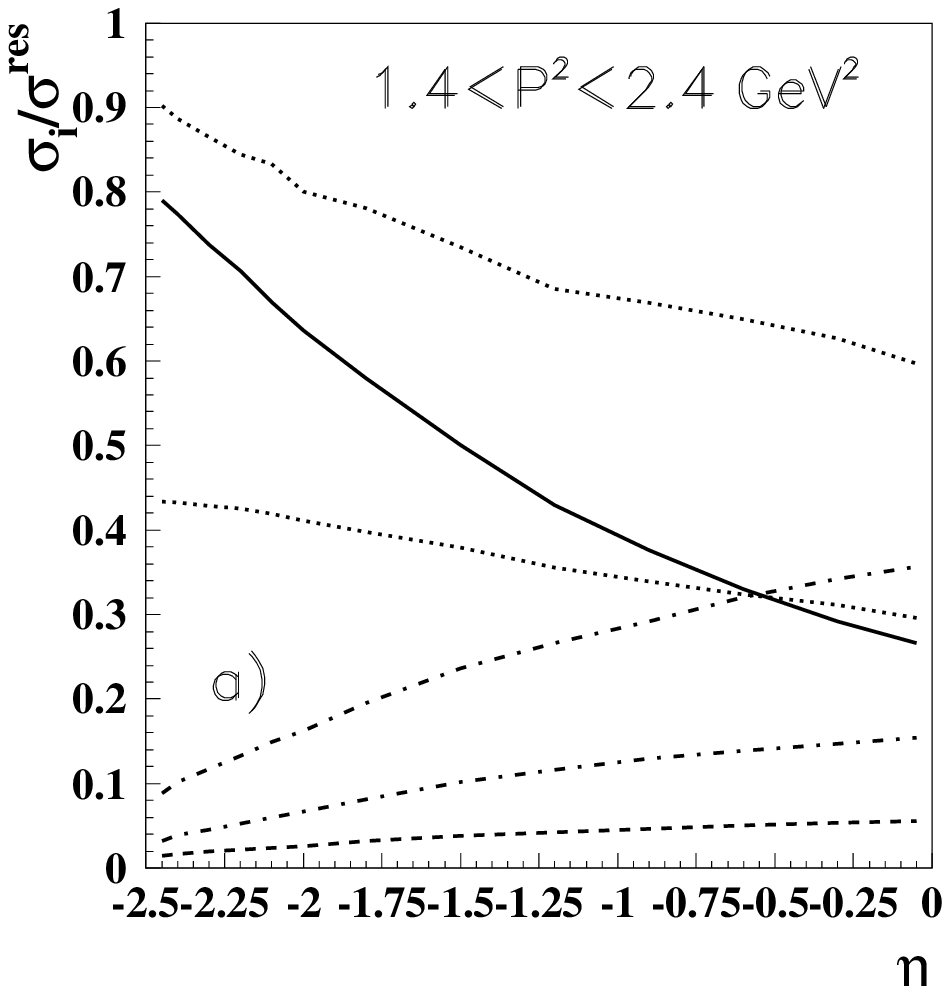,width=5.5cm}}
%\put(0,0){\epsfig{file=fractions2.eps,width=5.5cm}}
\put(55,0){\epsfig{file=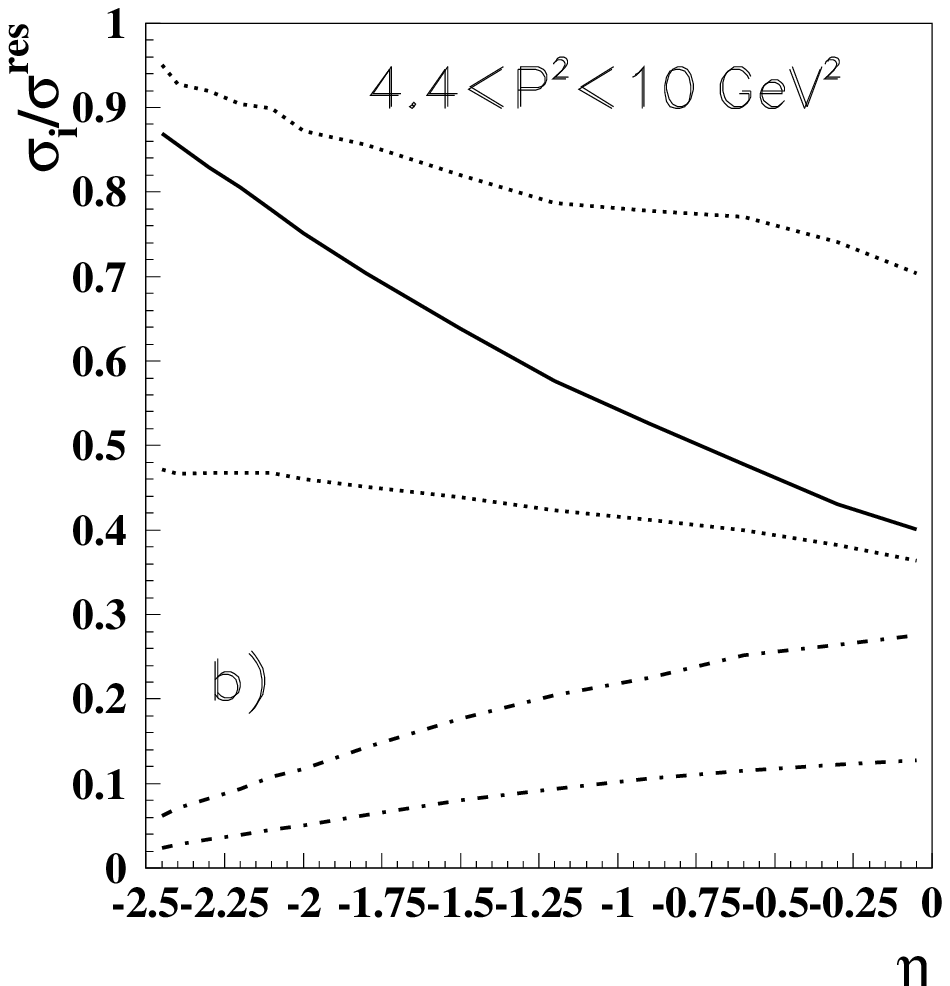,width=5.5cm}}
\put(110,0){\epsfig{file=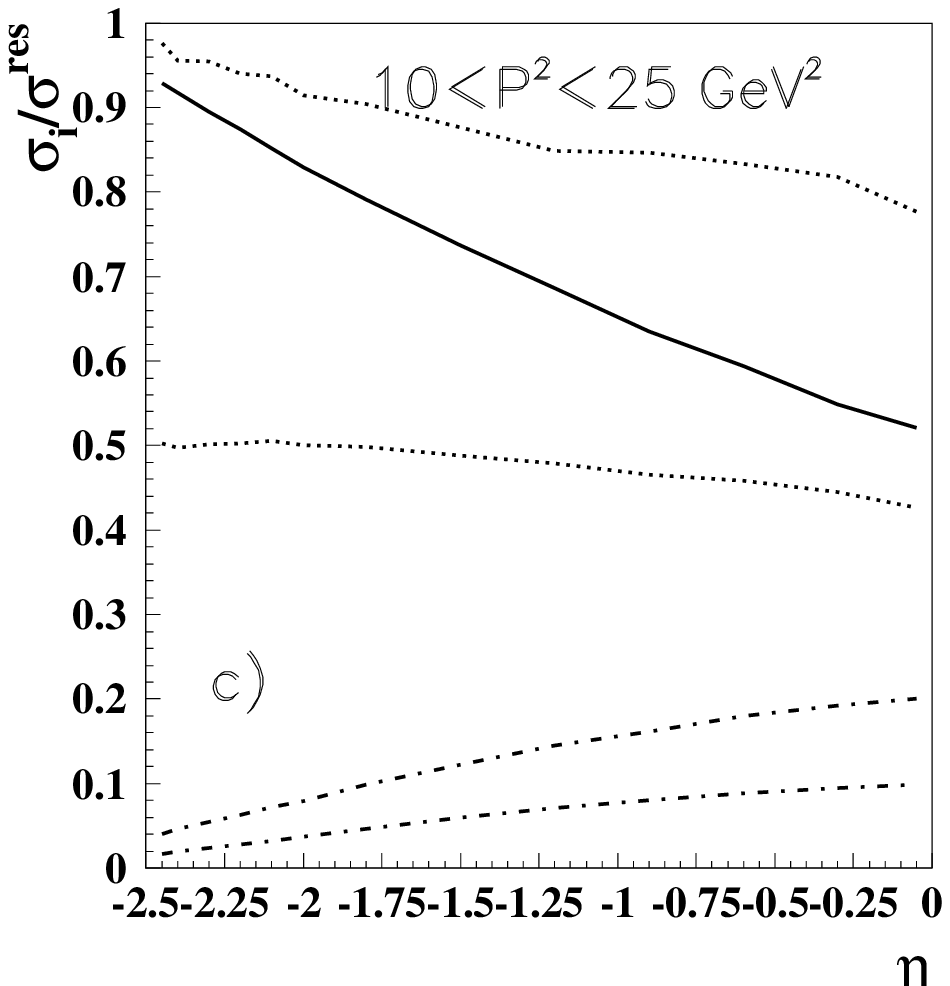,width=5.5cm}}
\end{picture}
\caption{Fractional contributions to
$\mathrm{d}\sigma^{\mathrm {res}}/\mathrm{d}\eta$.
Upper and lower dotted (dashed--dotted)
curves correspond to pointlike quarks (gluons) convoluted with
$O(\alpha_s^3)$ and $O(\alpha_s^2)$ partonic cross--sections.
The dashed curve in a) corresponds to the NLO VDM contribution.
The solid curves denote the ratio DIR$_{\mathrm{uns}}/$DIR$+$RES.}
\label{fractions}
\end{figure}

\begin{figure}\centering
\epsfig{file=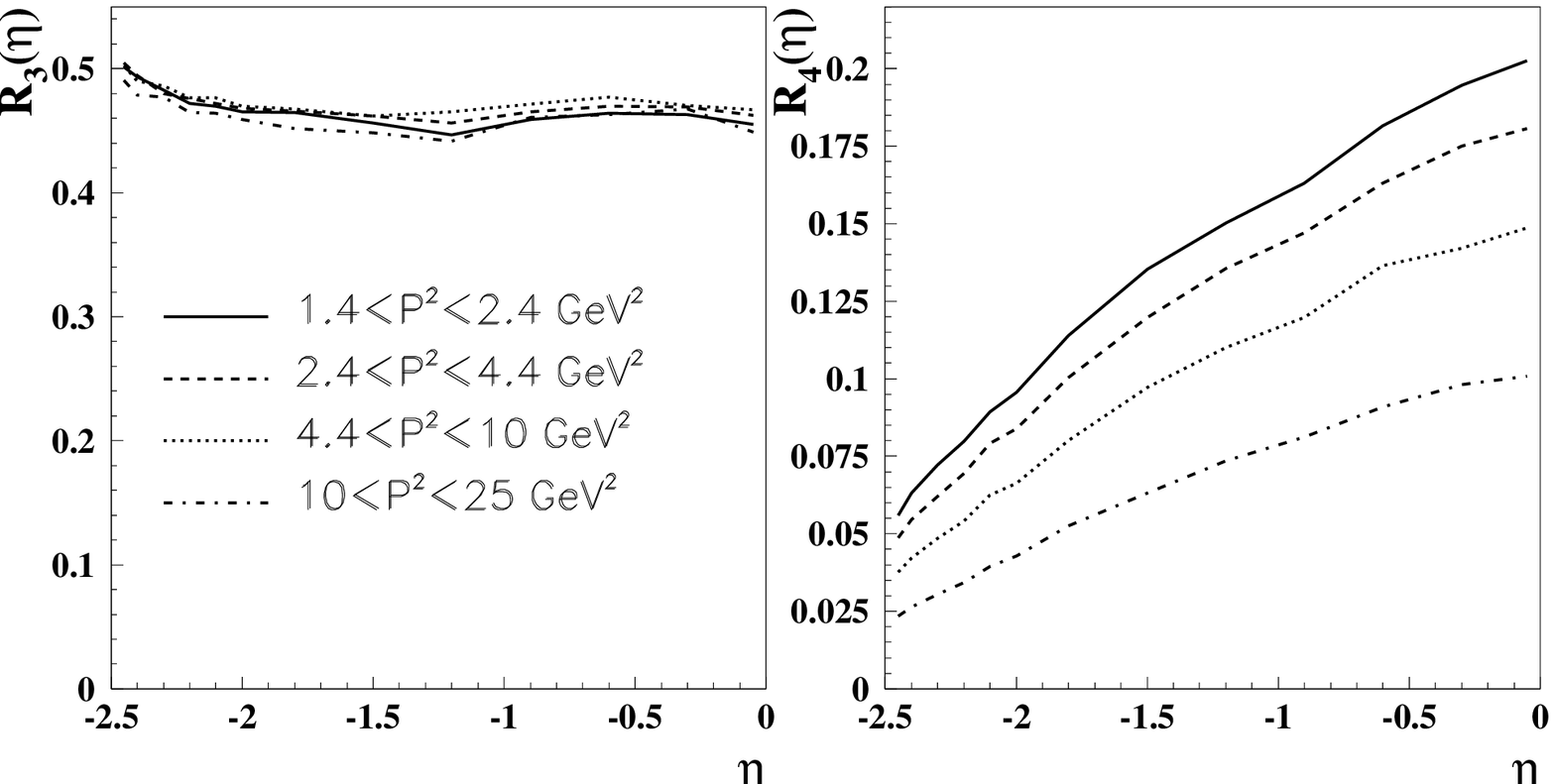,width=\textwidth}
\caption{Nontriviality fractions $R_3$ and $R_4$ as functions of
$\eta$ and $P^2$.}
\label{nontrivial}
\end{figure}

We start by plotting in Fig. \ref{win1} the distributions
${\mathrm {d}}\sigma/{\mathrm {d}}\eta$
and ${\mathrm {d}}\sigma/{\mathrm {d}}E_T$
in the first window
$1.4\le P^2\le 2.4$ GeV$^2$. All curves correspond to
$R_{\mathrm sep}=2$.
The difference between the solid and dashed curves is
significant in the whole range of $\eta$, but becomes truly
large close to the upper edge $\eta=0$, where the DIR$+$RES results
exceed the DIR$_{\mathrm {uns}}$ ones by a factor of about 3!
In ${\mathrm {d}}\sigma/{\mathrm {d}}E_T$ distributions this
difference comes predominantly from the region of $E_T$ close to
the lower cut--off $E_T^c+\Delta=7$ GeV.
Fig. \ref{win1} also shows that the scale dependence
is nonnegligible for both DIR$_{\mathrm{uns}}$ and DIR$+$RES results,
but does not invalidate the main conclusion drawn from this comparison.
Interestingly, the scale dependence is weaker for the DIR$+$RES
results than for the DIR$_{\mathrm uns}$ ones. Fig. \ref{drsep}
documents that the above results depend only very weakly on $R_{\mathrm
sep}$.

To track down the origins of the observed large differences
between DIR$+$RES and DIR$_{\mathrm {uns}}$ results we compare
in Fig. \ref{reseta1} the DIR$+$RES and DIR$_{\mathrm {uns}}$
results to the subtracted direct ones (denoted DIR).
The difference between the DIR$+$RES and DIR curves, giving the
resolved photon contribution ${\mathrm d}\sigma^{\mathrm
res}/{\mathrm d}\eta$, is further split into the following
contributions:
\begin{itemize}
\item VDM part of photonic PDF convoluted with complete NLO parton
level cross--sections (denoted NLO VDM).
\item Pointlike quarks and gluons convoluted with
$\sigma^{\mathrm{res}}(\alpha_s^2)$ and
$\sigma^{\mathrm{res}}(\alpha_s^3)$ parton level
cross--sections
\end{itemize}
displayed separately in the upper right plot of Fig. \ref{reseta1}.
The full NLO resolved photon contribution is given as the sum
\begin{equation}
\mathrm{NLO~VDM}+\sigma_q^{\mathrm{res}}(\alpha_s^2)
\otimes q^{\mathrm{PL}}+
\sigma_G^{\mathrm{res}}(\alpha_s^2)\otimes G^{\mathrm{PL}}+
\sigma_q^{\mathrm{res}}(\alpha_s^3)\otimes q^{\mathrm{PL}}+
\sigma_G^{\mathrm{res}}(\alpha_s^3)\otimes G^{\mathrm{PL}}.
\label{NLOsum}
\end{equation}
Fractional contributions of LO and NLO terms to
$\sigma^{\mathrm {res}}$ coming separately from pointlike
quarks and gluons
are plotted in Fig. \ref{fractions}a as functions of $\eta$.
Several conclusions can be drawn from Figs.
\ref{reseta1}--\ref{fractions}:
\begin{itemize}
\item The contribution of the VDM part of photonic PDF is very small
and perceptible only close to $\eta=0$.
Integrally it amounts to about 3\%. Using SaS2D
parameterizations would roughly double this number.
\item The inclusion of $\sigma_i^{\mathrm{res}}(\alpha_s^3)$
parton level
cross--sections in the resolved photon channel
is numerically very important throughout the range
$-2.5\le \eta\le 0$. Interestingly, the
$\sigma_i^{\mathrm{res}}(\alpha_s^3)$ results are close,
particularly for the pointlike quarks, to
the $\sigma_i^{\mathrm{res}}(\alpha_s^2)$ ones.
\item At both $\alpha_s^2$ and $\alpha_s^3$ orders
pointlike quarks dominate
${\mathrm d}\sigma^{\mathrm res}/{\mathrm d}\eta$
at large negative $\eta$, whereas
as $\eta\rightarrow 0$ the fraction of ${\mathrm d}\sigma^{\mathrm
res}/{\mathrm d}\eta$ coming from pointlike gluons increases
towards $40$\% at $\eta=0$.
\end{itemize}
We emphasize that pointlike gluons carry nontrivial information
already in convolutions with $\sigma^{\mathrm{res}}(\alpha_s^2)$
partonic
cross--sections because in unsubtracted direct calculations such
contributions would appear first at the order $\alpha\alpha_s^3$
\footnote{For
instance, the resolved photon diagram in Fig. \ref{nnlo}e would
come as part of the of evaluating the unsubtracted direct diagram
in Fig. \ref{nnlo}k.}.
The convolution of the dominant part of pointlike quarks
\footnote{That is, the one given by the splitting term
(\ref{splitterm}).}
with $\sigma^{\mathrm{res}}(\alpha_s^3)$
partonic cross--sections would be included
in direct unsubtracted calculations starting also at the order
$\alpha\alpha_s^3$, whereas for pointlike gluons this would require
evaluating the unsubtracted direct terms of even higher order
$\alpha\alpha_s^4$! For instance, the contribution of diagram in
Fig. \ref{nnlo}g would be included in the contribution of diagram
in Fig. \ref{nnlo}f. Similarly, the results of diagram in Fig.
\ref{nnlo}h would come as part of the results of evaluating the
diagram in Fig. \ref{nnlo}m.

In JETVIP the nontrivial aspects of taking into account the
$\sigma_i^{\mathrm{res}}(\alpha_s^3)$ resolved photon contributions
can be characterized
\footnote{Disregarding the VDM part of resolved contribution
which is tiny in our region of photon virtualities.} by the
``nontriviality fractions'' $R_3$ and $R_4$
\begin{equation}
R_3\equiv
\frac{q^{\mathrm {PL}}\otimes\sigma_q^{\mathrm {res}}(\alpha_s^3)+
      G^{\mathrm {PL}}\otimes\sigma_G^{\mathrm {res}}(\alpha_s^2)}
      {\sigma^{\mathrm {res}}},~~~
R_4\equiv
\frac{G^{\mathrm {PL}}\otimes\sigma_G^{\mathrm {res}}(\alpha_s^3)}
    {\sigma^{\mathrm {res}}},
\label{nontr}
\end{equation}
which quantify the fractions of $\sigma^{\mathrm {res}}$ that are
not included in NLO unsubtracted direct calculations. These
fractions are plotted as functions of $\eta$ and $P^2$ in Fig.
\ref{nontrivial}. Note that at $\eta=0$ almost 70\% of
$\sigma^{\mathrm {res}}$ comes from these origins. This fraction
rises even further in the region $\eta>0$, which, however, is
experimentally difficult to access.

So far we have discussed the situation in the first window of
photon virtuality, i.e. for $1.4\le P^2\le 2.4$ GeV$^2$. As $P^2$
increases the patterns of scale and $R_{\mathrm {sep}}$
dependencies change very little.
On the other hand, the fractions plotted in Fig.
\ref{fractions} and \ref{nontrivial} vary noticeably:
\begin{itemize}
\item The DIR$_{\mathrm {uns}}$ contributions represent an increasing
fractions of the DIR$+$RES results.
\item The relative contribution of pointlike gluons with respect to
pointlike quarks decreases.
\item The nontriviality factor $R_4$ (which comes entirely from
pointlike gluons) decreases, whereas $R_3$, which is dominated by
pointlike quarks and flat in $\eta$, is almost independent of
$P^2$.
\end{itemize}
All these features of JETVIP results reflect the fundamental fact
that as $P^2$ rises towards the factorizations scale $M^2\approx
E_T^2$ the higher order effects incorporated in pointlike parts of
photonic PDF vanish and consequently the unsubtracted direct
results approach the DIR$+$RES ones. The crucial point is that for
pointlike quarks and gluons this approach is governed by the ratio
$P^2/M^2$ appearing in the multiplicative factor $(1-P^2/M^2)$.
The nontrivial effects included in PDF of the virtual
photon will thus persist for arbitrarily large $P^2$, provided we
stay in the region where $P^2\ll M^2$.
Moreover, they are so large, that they should be visible already
in existing HERA data. Provided the basic ideas behind
the Schuler--Sj\"{o}strand
parameterizations of PDF of the virtual photon are correct, our
analysis shows that the calculations that do not introduce the
concept of virtual photon structure should significantly
undershoot the available HERA data on dijet production
in the kinematical region $1\lesssim P^2\ll E_T^2,~E_T\ge 5$ GeV,
$-2.5\le\eta\le 0$, in particular for $\eta\simeq 0$. Published 
as well preliminary
data discussed in \cite{phd,jarda} support this conjecture.

\section{Summary and conclusions}
We have analyzed the physical content of parton distribution
functions of the virtual photon within the framework formulated
by Schuler and Sj\"{o}strand, which provides physically motivated
separation of quark and gluon distribution functions into
their hadronic (VDM) and pointlike parts. We have shown that the
inherent ambiguity of this separation, numerically large for the
real photon, becomes phenomenologically largely irrelevant for
virtual photons with $P^2\gtrsim 2-3$ GeV$^2$. In this region quark
and gluon distribution functions of the virtual photon are
dominated by their (reasonably unique) pointlike parts, which have
clear physical origins. We have analyzed the nontrivial aspects of
these pointlike distribution functions and, in particular, pointed
out the role of pointlike gluons in leading order calculations of
jet cross--section at HERA.

The conclusions made within the framework of LO QCD have been
confirmed, and in a sense even strengthened, in our analysis of NLO
parton level calculations using JETVIP. We have found a significant
difference between JETVIP results in approaches with and without
the concept of virtual photon structure. While for the real photon
analogous difference is in part ascribed to the VDM part of
photonic PDF, for moderately virtual photons it comes almost
entirely from the pointlike parts of quark and gluon distribution
functions. Although their contributions are in principle contained
in higher order calculations which do not use the concept of PDF,
in practice this would require calculating at least
$\sigma(\alpha\alpha_s^3)$ and $\sigma(\alpha\alpha_s^4)$
unsubtracted direct
contributions. In the absence of such calculations the concept of
PDF of the virtual photon is therefore very useful
phenomenologically and, indeed, indispensable for satisfactory
description of existing data.

\vspace*{0.2cm}
\noindent
{\Large \bf Acknowledgment:}
We are grateful to J. Cvach, J. Field, Ch. Friberg, G. Kramer,
B. P\"{o}tter, I. Schienbein and A. Valk\'{a}rov\'{a}
for interesting discussions
concerning the structure and interactions of virtual photons and 
to B. P\"{o}tter for help in running JETVIP. This work was supported
in part by the Grant Agency of the Academy od Sciences of the Czech
Republic under grant No. A1010821.

\end{document}